\documentclass[aps,preprint,nofootinbib,preprintnumbers,eqsecnum,superscriptaddress]{revtex4}

\usepackage{amsmath} 
\usepackage{graphicx} 
\usepackage{amsthm}
\usepackage{amssymb} 
\usepackage{dsfont}
\usepackage{yfonts}
\usepackage{hyperref}
\usepackage{array,xcolor,graphicx}
\usepackage{booktabs,multirow}
\usepackage[utf8]{inputenc}
\usepackage{mathtools}

\usepackage{slashed}
\usepackage{caption}
\usepackage{stackrel}
\usepackage{cases}
\usepackage{tcolorbox}
\usepackage{etoolbox}
\patchcmd{\section}
  {\centering}
  {\raggedright}
  {}
  {}
\patchcmd{\subsection}
  {\centering}
  {\raggedright}
  {}
  {}

\usepackage[normalem]{ulem}
\usepackage{mathtools}

\usepackage[all]{xy}
\usepackage{tikz}
\usetikzlibrary{arrows.meta}

\hypersetup{colorlinks=true,linkcolor=blue,citecolor=red,urlcolor=blue, linktocpage=true}

\newcommand{\be}{\begin{equation}}
\newcommand{\ee}{\end{equation}}
\newcommand{\bea}{\begin{eqnarray}}
\newcommand{\eea}{\end{eqnarray}}

\newcommand{\bln}{\begin{align}}
\newcommand{\eln}{\end{align}}
\newcommand{\bst}{\begin{split}}
\newcommand{\est}{\end{split}}
\newcommand{\bi}{\begin{itemize}}
\newcommand{\ei}{\end{itemize}}
\newcommand{\ben}{\begin{enumerate}}
\newcommand{\een}{\end{enumerate}}

\def\le{\left}
\def\ri{\right}
\def\ha{{1\over 2}}

\def\al{{\alpha}}
\def\ket#1{|#1\rangle}

\def\vev#1{\langle#1\rangle}
\def\det{{\rm det}}

\def\Tr{\mathrm{Tr}}

\def\th{{\theta}}

\def\ep{{\epsilon}}

\newcommand{\p}{\partial}

\newcommand\ga{{\ensuremath{{\gamma}}}}

\newcommand\sig{\sigma}

\newcommand\lam{\lambda}
\newcommand\Lam{\Lambda}
\newcommand\om{\omega}

\def\lam{{\lambda}}

\def\eeq{\end{equation}}

\def\Tr{\mathop{\rm Tr}}

\newcommand\sL{{\ensuremath{{\mathcal L}}}}
\newcommand\sM{{\ensuremath{{\mathcal M}}}}

\newcommand\sO{{\ensuremath{{\mathcal O}}}}

\newcommand\bpsi{{\bar \psi}}

\def\th{{\theta}}

\numberwithin{equation}{section}

\begin{document}

\title{Jena lectures on generalized global symmetries: principles and applications}
\author{Nabil Iqbal} 
\email{nabil.iqbal@durham.ac.uk}
\affiliation{Centre for Particle Theory, Department of Mathematical Sciences, Durham University,
		South Road, Durham DH1 3LE, UK\\} 
\begin{abstract}
This is an elementary set of lectures on generalized global symmetries originally given at the Jena TPI School on QFT and Holography, designed to be accessible to the reader with a basic knowledge of quantum field theory. Topics covered include an introduction to higher-form symmetries with selected applications:  Abelian and non-Abelian gauge theories in the continuum and the lattice, statistical mechanical systems, the Adler-Bell-Jackiw anomaly from the point of view of non-invertible symmetry, and magnetohydrodynamics. Mathematical formalism is kept to a minimum and an emphasis is placed on understanding the global symmetry structure present in well-known models.
\end{abstract} 
\maketitle

\tableofcontents

\newpage

\section{Introduction}

One of our most basic tools in the quantitative understanding of nature is the idea of {\it symmetry.} Any reader who has opened these lecture notes probably does not need to be convinced that symmetries are important. This hypothetical reader might, however, be surprised to know that in recent years the very idea of symmetry in quantum field theory has undergone something of a renaissance. As we will discuss in detail below, it turns out that the symmetries that one learns about in an introductory physics course are actually only a {\it very special case} of a much richer structure. This richer structure is sometimes called that of {\it generalized global symmetries} \cite{Gaiotto:2014kfa}. 

This new set of ideas captures very basic physical principles. It is not complicated or overly formal (at least, not more so than most quantum field theory already is), and it seems to the author that it could very easily be taught in elementary quantum field theory courses. It is the goal of these lecture notes -- based on a set of lectures first given at the Jena TPI School on QFT and Holography -- to present the material at such an introductory level, and thus to equip a reader who already has a working knowledge of quantum field theory with a similar working toolkit of these new symmetries. 

The field is vast and growing rapidly, and these notes do not attempt to provide an authoritative overview. Instead they necessarily discuss a rather narrow and somewhat biased sampling of the ideas, centered around explaining how these generalized global symmetries exist already in many of the simplest and most familiar systems -- e.g. ordinary electrodynamics, or pure Yang-Mills theory, or even the 2d Ising model -- that theoretical physicists study. Several other recent reviews discussing different aspects of generalized symmetries include \cite{Cordova:2022ruw,Brennan:2023mmt,Gomes:2023ahz,Shao:2023gho,Schafer-Nameki:2023jdn,Bhardwaj:2023kri}. Our discussion is philosophically aligned with the point of view of \cite{McGreevy:2022oyu}. Finally, though we follow a viewpoint from quantum field theory that was formalized in \cite{Gaiotto:2014kfa}, very similar ideas have been discussed in the condensed-matter literature earlier \cite{Nussinov:2006iva,Nussinov:2009zz}\footnote{See also \cite{Batista:2004sc,Nussinov:2011mz}.}. 

\subsection{Motivation: what is gauge theory?} \label{sec:motivation} 

Before beginning, let us start with a question. In most elementary treatments of field theory, the concept of {\it gauge theory} plays a central role. 

What, then, is a gauge theory? 

At first, this seems to be a silly question: clearly, a gauge theory is a theory with a gauge field. For example, in three spacetime dimensions, we could study free electrodynamics, with action
\be
S_{\mathrm{gauge\;theory}}[A] = \frac{1}{4g^2} \int d^3x F_{ij} F^{ij} \qquad F_{ij} = \p_{i} A_{j} - \p_{j} A_{i} \label{maxwell4d} 
\ee
This is obviously a gauge theory. 

However, in three dimensions, it is a well-known fact that one can {\it dualize} it, i.e. rewrite this gauge theory in terms of a massless scalar field $\th$ using the following change of variables:
\be
\frac{1}{g^2} F_{ij} = \ep_{ijk} \p^k \theta \label{dualmap} 
\ee
and then write down an action describing the system in terms of the new scalar field as
\be
S_{\mathrm{gauge\;theory?}}[\th] = \frac{g^2}{2} \int d^3x\;(\p_i \theta)(\p^i \theta) \label{scalar3d}  \ . 
\ee
The theory described by \eqref{scalar3d} is precisely equivalent to \eqref{maxwell4d}, but simply expressed in different variables\footnote{This is the mapping between symmetry-broken phases of the two quantum field theories -- a charged scalar field and the Abelian-Higgs model -- which are related by {\it particle-vortex duality} \cite{Peskin:1977kp,dasgupta1981phase}; see e.g. Chapter 8 of \cite{tong2018gauge} for a review.}.

Now it seems that it would be ridiculous to claim that a free massless scalar in three dimensions is a gauge theory; yet if \eqref{scalar3d} is truly the same as \eqref{maxwell4d}, then apparently it is. 

This problem is not restricted to this example, and actually sits at the heart of many of the most fascinating phenomena in theoretical physics. Some other systems which are not obviously gauge theories -- but which nevertheless can be {\it usefully} described by Lagrangians involving gauge fields -- include systems as disparate as (the right density of) electrons in a magnetic field (i.e. the fractional quantum hall effect), or quantum gravity on AdS$_5 \times S^5$ (i.e. the AdS/CFT correspondence \cite{Maldacena:1997re}). The issue here is that a gauge symmetry is not really a symmetry at all -- it is more precisely stated as a {\it redundancy}, a feature of the variables that we choose to describe the system, rather than an intrinsic property of the system itself. 

But if gauge fields can come and go in this way when we change the variables that we use, then what is the {\it point} of gauge theory? A potential answer to this question is that gauge theory is intimately associated with describing {\it extended structures} -- e.g. in the most familiar case of electrodynamics, the extended objects in question are electric and magnetic field lines. We should then seek an organizing framework built around the existence of such extended objects. Generalized global symmetries -- as we will see in detail below -- provide precisely such a framework. 

\begin{tcolorbox}

{\bf Exercise}

Prove that the quantum theories defined by \eqref{maxwell4d} and \eqref{scalar3d} are equivalent, by placing the action \eqref{maxwell4d} in a path integral and performing an appropriate change of variables.

\end{tcolorbox} 
 
\section{Ordinary global symmetries, phases of matter, and the Landau paradigm}
These lectures will be about {\it new} sorts of global symmetries. Let's start by carefully understanding ordinary global symmetries. 
\subsection{Ordinary global symmetries}

To orient ourselves, let's consider a field theory with a $U(1)$ global symmetry. For much of what we will discuss in this section, you can consider any field theory you like, but for concreteness let us consider a charged scalar field:
\be
S[\phi] = \int d^{d}x \le(\p \phi^{\dagger} \p \phi + V(\phi^{\dagger}\phi)\ri)  \label{compscalac} 
\ee
We will assume that the potential has at least a mass term, and possibly some self interactions:
\be
V(\phi^{\dagger}\phi) = m^2 \phi^{\dagger} \phi + \frac{\lambda}{4}(\phi^{\dagger}\phi)^2 + \cdots \label{pot}
\ee
In almost all of these lectures we will work in Euclidean spacetime with $d$ dimensions, where $d$ will take various interesting values as we proceed. 

This is a classical action, but we can make it into a quantum theory by inserting its exponential into a path integral:
\be
Z = \int [d\phi] \exp(-S[\phi,\phi^{\dagger}])
\ee
Now we know very well that this theory has a $U(1)$ global symmetry. What does it {\it act on?} This is a question with an obvious answer: it acts on the fundamental degree of freedom, the $\phi$ field, as
\be
\phi(x) \to \phi'(x) = e^{i\alpha} \phi(x) 
\ee
where $e^{i\alpha} \in U(1)$ is a space-time independent phase, and clearly the action is invariant. $\phi(x)$ is a local operator. Now through the usual Noether procedure we can construct a current which has {\it one index} for this symmetry:
\be
j^{\mu}(x) = i(\phi^{\dagger} \p^{\mu} \phi - \phi \p^{\mu} \phi^{\dagger})
\ee
The current is conserved on-shell. Classically, this just means that if $\phi$ satisfies the equations of motion we have 
\be
\p_{\mu} j^{\mu} = 0 \ . \label{currcons} 
\ee
 This classical conservation equation is somewhat modified in the quantum theory. If we apply Noether's theorem inside the path integral, then (as explained in any introductory quantum field theory textbook), we find an expression of the form: 
\be
\langle \p_{\mu} j^{\mu}(x) \phi(y) \cdots \rangle = i \delta^{(d)}(x-y) \langle \phi(y) \cdots \rangle
\ee
In other words: the current is conserved {\it up to delta function terms at the location of the insertion of charged operators}. This is the {\it Ward identity}, and one should view this expression as telling us what it means to be charged operator. 

\begin{figure}[h]
\begin{center}
\includegraphics[scale=0.5]{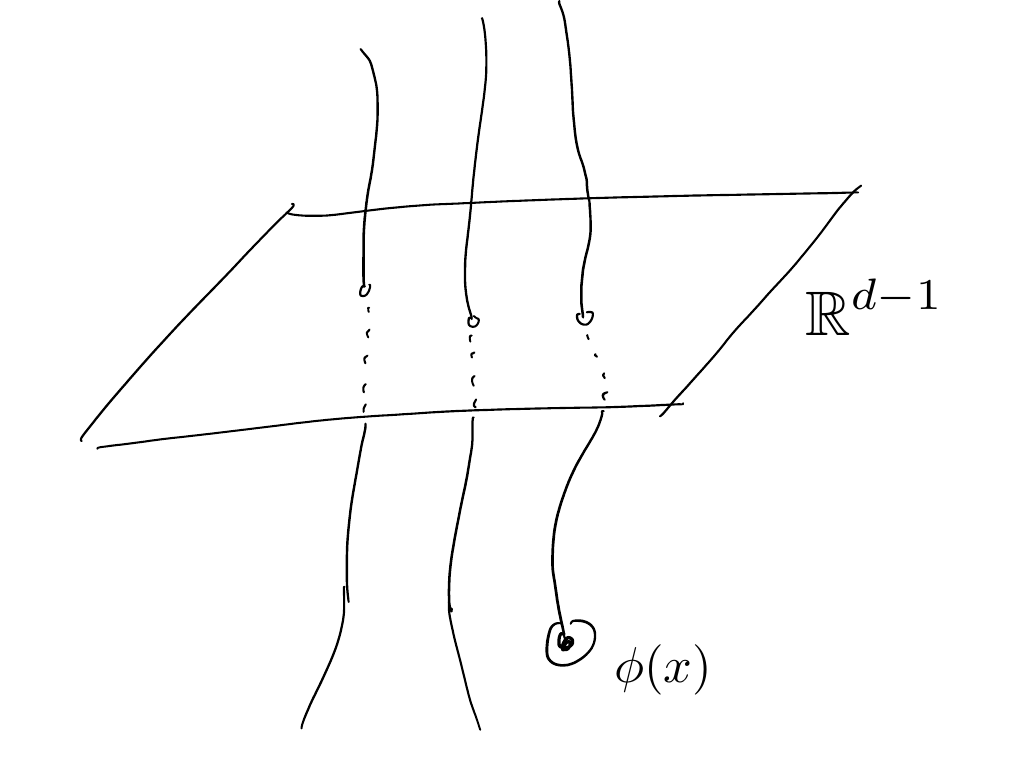}
\end{center}
\caption{Particle world-lines can't end: no matter at what time-slice we evaluate $Q(t)$ on we get the same number.}
\label{fig:0formparticles}
\end{figure}

Finally, global symmetries allow us to define a conserved charge. Normally we do this in Lorentzian signature in the following way:
\be
Q(t) = \int_{\mathbb{R}^{d-1}} d^{d-1}x j^0(t,\vec{x}) \label{ordQdef} 
\ee
where we do the integral over a time-slice $\mathbb{R}^{d-1}$. Let's just verify why this is conserved:
\be
\frac{d}{dt} Q(t) = \int_{\mathbb{R}^{d-1}} d^{d-1}x \p_{t} j^{0}(t,\vec{x}) = -\int_{\mathbb{R}^{d-1}} d^{d-1}x \p_{i} j^{i}(t,\vec{x}) = \mbox{boundary term} = 0
\ee
Let us draw a picture of this equation in Figure \ref{fig:0formparticles}: imagine that there are some particles sitting around. They trace out worldlines in space-time. The world-lines can't end, because the particle number is conserved. This integral over the time-slice intersects all of the world lines -- you can move it up and down in time and get the same answer, which is just the count of the particle number. A key point here is that {\bf ordinary symmetries are associated with counting particles.} 

Note also that the local field $\phi(x)$ is the thing that creates a particle or anti-particle. Another point is {\bf the charged object creates the thing that the charge counts}. 

One more piece of formalism is that it is often useful to couple in an external gauge field $a_{\mu}$ to the current $j_{\mu}$. Here this coupling is implemented by demanding that the system is invariant under the combined transformation:
\be
a \to a + d\Lambda \qquad \phi \to \phi e^{i\Lambda} \label{u1extsource} 
\ee
where now $\Lambda(x)$ is a function of spacetime. To do this, we have to replace the normal derivative $\p_{\mu}\phi$ with the gauge-covariant one $D_{\mu}\phi = \p_{\mu}\phi - i a_{\mu} \phi$, and we can consider the action:
\be
S[\phi; a] =  \int d^{d}x \le((D \phi)^{\dagger} D \phi +  V(\phi^{\dagger}\phi)\ri) \qquad Z[a] = \int [d\phi] \exp(-S[\phi; a])
\ee
Note that $a$ is a fixed external field, and the invariances above mean that the path integral is invariant under external gauge transformations of $a$:
\be
Z[a + d\Lambda] = Z[a]
\ee
Furthermore, we can compute the current by constructing $j^{\mu}(x) = \frac{\delta Z[a]}{\delta a_{\mu}(x)}$.

Now, what are global symmetries good for? 

\subsection{The old Landau paradigm}
One very important use of global symmetries is to classify the {\it phases of matter}. Note that the theory described above has two different phases, which we now study in turn: 

\begin{enumerate}
\item {\bf The Unbroken Phase}. Here we have $m^2 > 0$, and we have a unique vacuum in which the field $\phi$ is basically zero, i.e. $\langle \phi(x) \rangle = 0$. Because this vacuum is invariant under the $U(1)$ symmetry, we say that the $U(1)$ symmetry is {\bf unbroken.} 

In this phase we can imagine doing a calculation of the following two-point function:
\be
\langle \phi^{\dagger}(x)\phi(0)\rangle \sim \exp(-m|x|) \ . \label{expdecay} 
\ee
If we turn off the interactions, this is again a standard computation in elementary field theory. Physically, the reason that this decays exponentially in space is that $\phi(0)$ creates a particle at the origin which must then survive until it is annihilated at $\phi^{\dagger}(x)$. This costs action because the particle is heavy, and thus the exponential decay. 

\item {\bf The Spontaneously Broken Phase}. Here we have $m^2 < 0$, and the potential \eqref{pot} has a family of minima at $\phi^{\dagger}\phi = \frac{\lam}{2 m^2} \equiv v^2$. Thus the vacuum will have a nonzero value of $\langle \phi \rangle = v$.  In this phase we say that the particles have condensed, and the $U(1)$ symmetry is {\bf spontaneously broken}. 

We can parametrize the interesting degree of freedom here by writing
\be
\phi(x) = v(x) e^{i\theta(x)}
\ee
describing the modulus and phase of the field respectively.
What can we say about the dynamics in the broken phase? Now plugging this into \eqref{compscalac} we see that the field $v(x)$ acquires a mass which pins it at the minimum of the potential. Thus at low energies we may ignore it. However the phase field $\theta(x)$ is massless, so the leading low-energy action is 
\be
S = \int d^dx\;v^2 (\p \theta)^2 + \cdots \label{0formGoldstone} 
\ee
$\theta$ is the Goldstone mode of the symmetry. Its masslessness is much stronger than the classical argument we just gave: Goldstone's theorem states non-perturbatively that if there is a spontaneously broken symmetry (as defined by $\vev{\phi} \neq 0$) then there is necessarily a gapless mode in the spectrum. We will prove this again shortly as a warm-up for fancier things. 

Let us consider another way to describe the broken symmetry. Consider computing 
\be
\lim_{|x| \to \infty}\langle \phi^{\dagger}(x) \phi(0) \rangle \sim \langle \phi^{\dagger}(x) \rangle \langle \phi(0) \rangle = v^2 \label{odlro} 
\ee
Here the first equality follows from cluster decomposition, i.e. the fact that things far away should be independent. Thus an equivalent way to state that the symmetry is spontaneously broken is to note that the two-point function of the charged field saturates and factorizes, becoming independent of the separation, as contrasted with the exponential decay shown in \eqref{expdecay}. In this form this is sometimes called {\it off-diagonal long-range order}. 

We say that $\phi$ is the {\bf order parameter} for the transition, because it behaves so differently in the two phases. 
\end{enumerate} 
Now, basically when you have a symmetry, these are the two phases that you can have: unbroken and  spontaneously broken. You can also ask what happens right between the two: there you have a {\it phase transition}: 
\begin{enumerate}
\setcounter{enumi}{2}
\item {\bf Phase Transition}: in the model above, the boundary between the two phases is at $m^2 = 0$. It turns out that at such a continuous transition we then have a {\it conformal field theory} which is characterized by the symmetry group in question (in this case $U(1)$) and the number of dimensions. It seems reasonable to describe the transition by the action we wrote down earlier, but setting $m^2 = 0$: 
\be
S = \int d^{d}x \le(\p \phi^{\dagger} \p \phi + \frac{\lambda}{4}(\phi^{\dagger}\phi)^2 + \cdots \ri)  \label{CFT} 
\ee
If we now compute the correlation function of the charged operator we will find something which is {\it in between} the two behaviors (i.e. constant and exponential) shown above: it is a power law. 
\be
\langle \phi^{\dagger}(x) \phi(0) \rangle \sim \frac{1}{|x|^{2\Delta}}\, ,
\ee
where $\Delta$ is a number that is interesting to calculate -- it is called an {\it operator dimension} in CFT or a {\it critical exponent} in statistical physics. As it turns out, for $d < 4$ this number depends on the interactions in an important way.  
\end{enumerate} 
Note that from this point of view, the reason why a CFT exists is to describe the universal physics present at the phase transition between two distinct phases. 

Now the point of all of this is that it gives us a way to describe phases of matter. A phase of matter is a classification of the long-distance physics (as captured e.g. by the correlation functions \eqref{expdecay} and \eqref{odlro}), and in many cases it is determined completely by how the symmetries are realized.

Let us study some more examples: 
\ben
\item {\bf $U(1)$ superfluid:} if we imagine putting many atoms in a box, then they have a conserved particle number enforced by a $U(1)$ symmetry much like the one we just described\footnote{The effective theory is somewhat different because the system is non-relativistic, and the background value of the charge density $\vev{j^0}$ is nonzero.}. Here if the symmetry is {\it unbroken}, then we call it the normal phase. On the other hand, if you cool the system down enough (see e.g. \cite{ketterle1999experimental} for a discussion of the experimental situation), then the effective potential for the ``field that creates an atom'' develops a minimum away from the origin, and this field condenses, breaking the $U(1)$ number symmetry. This is now called a {\it superfluid}. 
\item {\bf Magnetization:} the simplest way to think about this is to imagine a bunch of spins $\vec{s}$ that are fixed on the sites of a lattice. There can be a non-Abelian rotational\footnote{You may be worried that in real life this should be a spacetime symmetry and not an internal symmetry. This is correct and manifests in spin-orbit couplings which can couple the spin to the orbital angular momentum of atoms. Often this can be neglected, or we work on explicit lattice systems where it is absent from the construction.} symmetry that acts on the spins as $\vec{s} \to R \vec{s}$. We often consider very simplified settings -- e.g. you could imagine a simpler symmetry group $\mathbb{Z}_2$ that simply acts on a scalar spin $s$ as $s \to \pm s$, which defines what is called the {\it Ising model} (we will review this model in Section \ref{sec:ising}).  If this rotational symmetry is unbroken, then the system is called a {\it paramagnet}, and if it spontaneously broken -- i.e. if all the spins are lined up -- we call it a {\it ferromagnet}. 
\item {\bf Liquids and solids:} imagine have a collection of atoms moving freely; sometimes they can stop moving and form a crystal, in which case {\it translational symmetry} is spontaneously broken down to some discrete subgroup (e.g. for a square lattice it would be $\mathbb{Z}^d$). A {\it solid} is thus a phase where translational symmetry is spontaneously broken, and a {\it liquid} is a phase where translational symmetry is unbroken. 

What about the difference between liquid and gas? These have the same symmetries, and thus do not {\it actually} define different phases -- this means that we can connect them continuously in a phase diagram, as is the case e.g. for \href{https://en.wikipedia.org/wiki/Phase\_diagram}{water}. 
\een 

A further point here is that the transitions between phases can be continuous or discontinuous (i.e. first order). If they are first order there is not much more we can say; but if they are continuous then we can write down an effective theory involving a coarse-grained order parameter to describe the transition basically as we did in \eqref{CFT}.

These ideas -- that the phases of matter are classified by patterns of spontaneously broken and unbroken symmetry, and that the transitions between them can be described by the symmetries alone -- are called the {\bf Landau paradigm} of the phases of matter. It is extremely successful and is one of the cornerstones of our understanding of nature. 

It also seems that it doesn't always work. 

\subsection{Physics beyond the Landau paradigm} 
Here are some phases which are outside the simplest Landau paradigm:
\ben
\item {\bf Gauge theory:} it seems that gauge theories can come in different phases. For $U(1)$ gauge theory in the continuum we have the Coulomb and Higgs phases, which seem different. What is the order parameter? There is no local order parameter for this, and similarly for non-Abelian gauge theory and confinement.  
\item {\bf Topological order:} sometimes there are phases of matter which interact in an interesting manner with the topology of the space where they are defined -- e.g. if you place a density of electrons in a strong magnetic field, you get the {\it fractional quantum Hall effect}, which basically says that the electron somehow splits into three smaller objects with fractional charge. These three pieces of the electron do interesting things if you move them around each other. Fascinatingly, the ground state degeneracy of this  phase is different if you put it on a torus or a sphere. 
\item {\bf Gravity:} gravity can be thought of as a phase of matter with a gaplessly dispersing mode. It does not appear to fit into the conventional Landau paradigm; we will not discuss this in these lectures, but some recent work in this direction appears in \cite{Hinterbichler:2022agn, Benedetti:2021lxj, Cheung:2024ypq}.
%\item {\bf Fermi (and non-Fermi) liquids} 
%\item {\bf Deconfined quantum criticality}: \cite{Senthil:2023vqd}. 
\een 

The goal of these lectures is to generalize our understanding of global symmetries in two directions: {\bf higher-form} global symmetries, and {\bf non-invertible} global symmetries. We will see that once we do this, we will actually be able to significantly extend the Landau paradigm in a way that encompasses some of the phases above. We stress that there are many {\it other} ways to go beyond the Landau paradigm: for an overview on how generalized symmetries interact with them, see \cite{McGreevy:2022oyu}. 

\subsection{Symmetries as topological operators} 
To do this, we first need to obtain a refined understanding of ordinary symmetries. Note that the description above made heavy use of the fundamental field $\phi(x)$: but from a modern point of view, in an age of duality where the notion of ``fundamental'' has little meaning, this seems quaint. We seek a notion of global symmetry that does not require any particular choice of fundamental fields, and works in a duality-invariant way. 

So, let us take a step back: if we have a symmetry, we have a conserved current:
\be
\p_{\mu} j^{\mu} = 0 \qquad d\star j = 0
\ee
Here we will start using a differential forms notation. Now let's consider the following operation: consider a codimension-$1$ manifold $\sM_{d-1}$ and form the following charge integral:
\be
Q(\sM_{d-1}) = \int_{\sM_{d-1}} d^{d-1}x \sqrt{h} \;n_{\mu} j^{\mu} = \int_{\sM_{d-1}} \star j \label{Qdef} 
\ee
This appears to be a function of the manifold $\sM_{d-1}$ -- however because the current $j^{\mu}$ is conserved, it is actually independent under small wiggles of $\sM_{d-1}$: it is a {\bf topological defect} in the theory. The notion of topological defect will appear multiple times, so here it is good to make an informal definition:

{\bf A topological defect on some submanifold $\sM$ is something that does not change under small deformations of $\sM$}. 

Note that it is exactly the same sort of integral that we do when constructing a usual conserved charge, except that we now take $\sM_{d-1}$ to be a possibly wiggly manifold rather than a flat time slice $\mathbb{R}^{d-1}$. 

\begin{figure}[h]
\begin{center}
\includegraphics[scale=0.5]{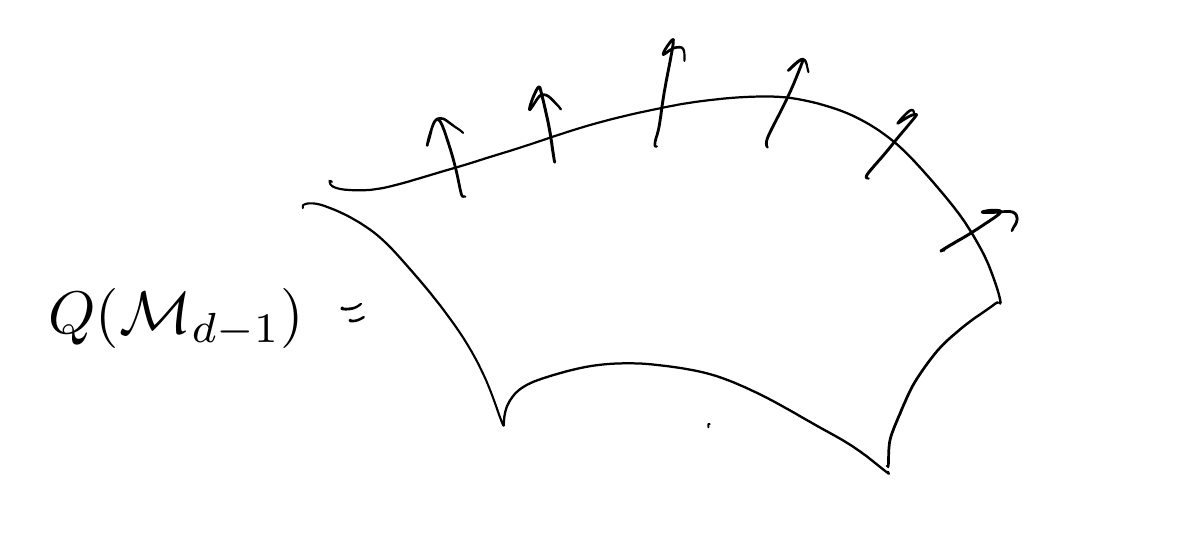}
\end{center}
\caption{Integrating divergenceless current $\star j$ on a codimension-1 manifold $\sM_{d-1}$.}
\label{fig:0formQmanifold}
\end{figure}

Thus whenever we have a normal symmetry, we obtain a {\bf codimension-$1$ topological defect}. This simple idea now supports many generalizations, which we will now explore. 

\begin{figure}[h]
\begin{center}
\includegraphics[scale=0.5]{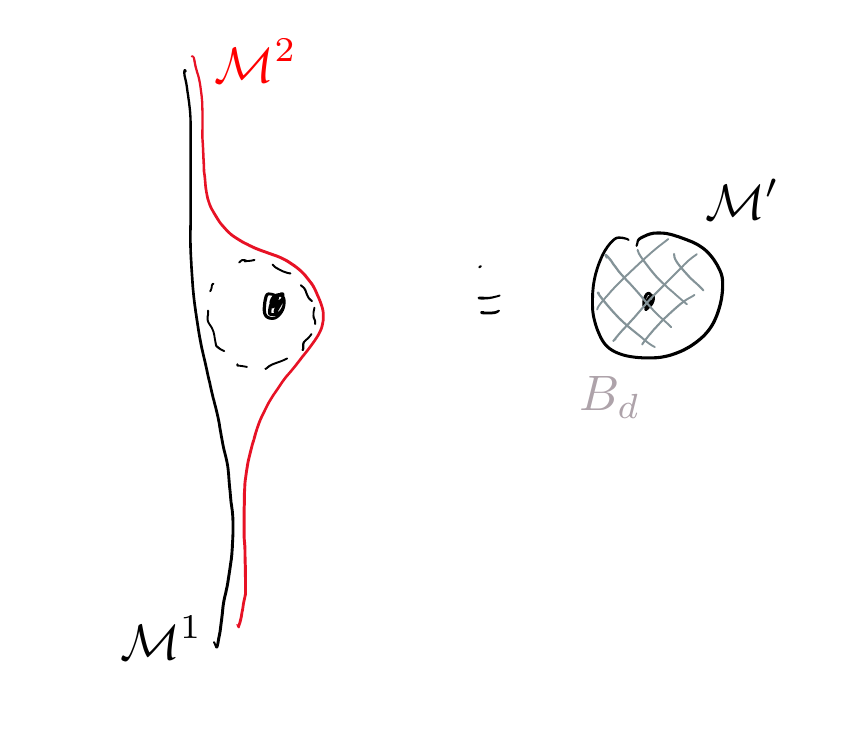}
\end{center}
\caption{Deforming $\sM^1$ to $\sM^2$ receives a contribution only from the insertion of the local operator $\sO(x)$.}
\label{fig:deformM1toM2}
\end{figure}

What happens when you consider the interaction of this charged defect with a charged operator $\sO(x)$? In the presence of a charged operator, we know from the Ward identity that the current is not conserved, i.e. we have
\be
\p_{\mu} j^{\mu}(x) \sO(y)  = i \delta^{(d)}(x-y) \sO(y)
\ee
(where it is understood that this expression holds in the path integral). Thus if we try to drag the topological defect $\sM_{d-1}$ through the insertion of a charged local operator, it not topological anymore; instead the operator picks up a phase, i.e. as in Figure \ref{fig:deformM1toM2}
\begin{align}
\left(Q(\sM_{d-1}^1)- Q(\sM_{d-1}^2)\right)\sO(x)  & = Q(\sM'_{d-1})\sO(x)  =\oint_{\sM'_{d-1}} \star j \sO(x)  \nonumber \\ & =  \int_{B_{d}} (d \star j) \sO(x) = i\int_{B_d} \delta(x-y) \sO(x) = i \sO(x) \label{movingthrough} 
\end{align}
where above $B_d$ ``fills in'' $\sM'_{d-1}$, i.e. $\p B_{d} = \sM'_{d-1}$. 
There's another way to say this. If we formulate the charge operator on an $S^{d-1}$ that wraps the operator and collapse it onto the operator as a point, we find:
\be
Q(S^{d-1}) \sO(x) = i \sO(x) \label{infu1} 
\ee
It is also convenient to consider the topological defect that constructs a {\it finite} charge rotation by an angle $\alpha$. As expected this is the exponential of the infinitesimal charge rotation, so we have:
\be
U_{\al}(\sM_{d-1}) \equiv \exp(i \al Q(\sM_{d-1})) \qquad U_{\al}(S^{d-1})\sO(x) = e^{i\alpha} \sO(x) \label{u1s2} 
\ee 
i.e. when we collapse the charge generator onto the local operator it acts as a linear phase rotation on the field. 

\begin{figure}[h]
\begin{center}
\includegraphics[scale=0.6]{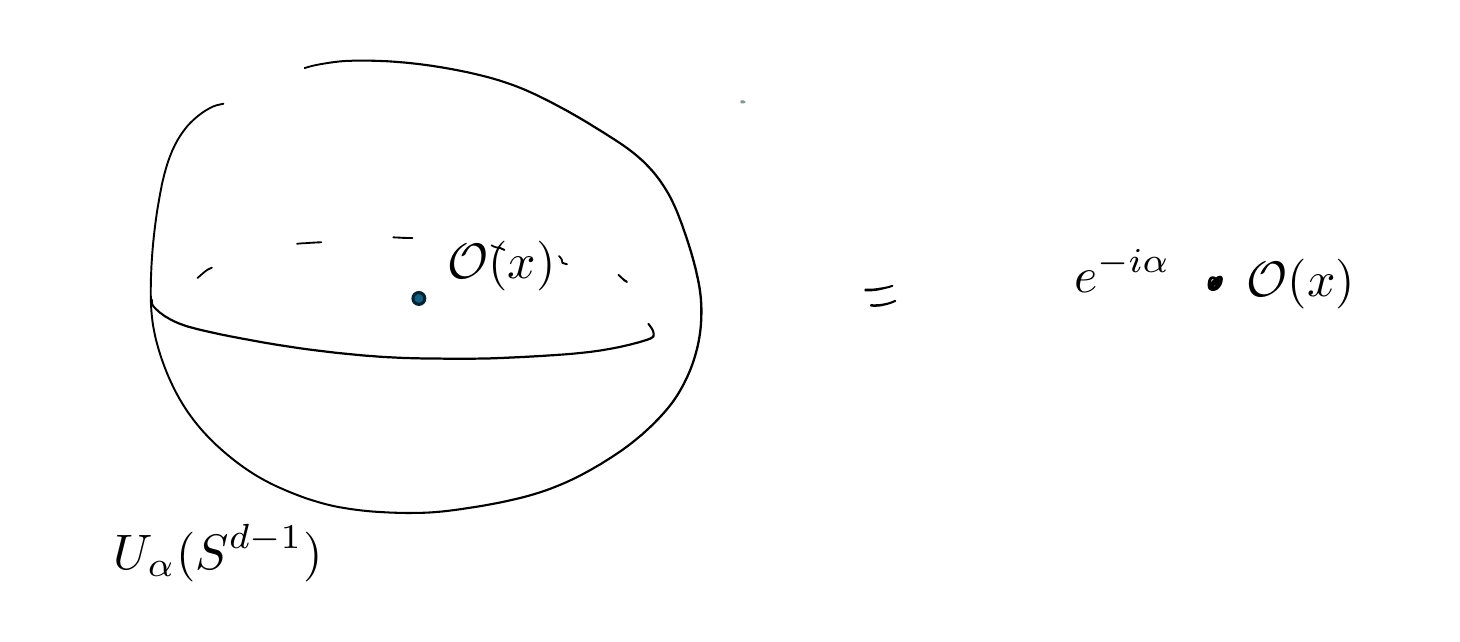}
\end{center}
\caption{Collapsing the charge generator $U_{\al}(S^{d-1})$ onto $\sO(x)$ acts as a phase rotation.}
\label{fig:collapseS0form}
\end{figure}

Note also that if you imagine {\it composing} two of these charge operators by forming them on parallel manifolds and fusing them together we find:
\be
U_{\al}(\sM_{d-1}) U_{\beta}(\sM_{d-1}) = U_{\alpha + \beta}(\sM_{d-1})
\ee
Here the last expression $\al + \beta$ should be understood in the spirit of group composition of the two $U(1)$ group elements $e^{i\al}$ and $e^{i\beta}$. 

\begin{figure}[h]
\begin{center}
\includegraphics[scale=0.5]{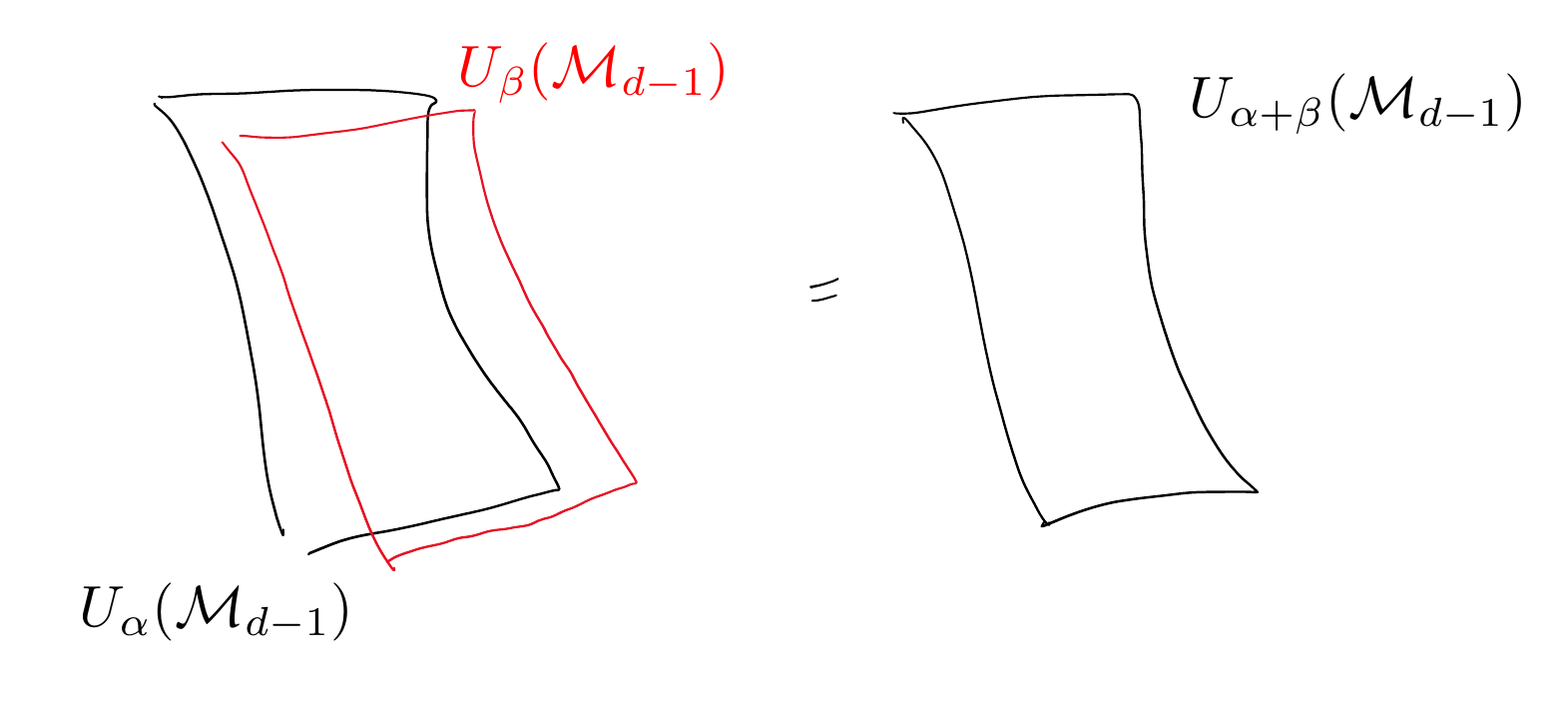}
\end{center}
\caption{Fusing together two charge defects results $U_{\al}$ and $U_{\beta}$ results in a combined charge defect $U_{\al+\beta}$.}
\label{fig:collapseS0form}
\end{figure}

We may now generalize this somewhat to an arbitrary non-Abelian global symmetry. Consider a symmetry group $G$, and imagine that we have fields $\sO_{a}$ that transform in some representation $r$ under $G$, so that for every $g \in G$ there are matrices $V^{a}_b(g)$ that provide a linear action of the group, so
\be
V^{a}_b(g) V^{b}_{c}(g') = V^{a}_c(gg')
\ee

The analogue of the above is now that for every element $g$ of the symmetry group $G$ we have topological defect operators $U_{g}(\sM_{d-1})$. They can be composed in the obvious way as follows:
\be
U_{g}(\sM_{d-1}) U_{g'}(\sM_{d-1}) = U_{g g'}(\sM_{d-1}) \label{groupcomp} 
\ee
To understand the action on local operators $\sO_{a}(x)$, we imagine forming a defect on an $S^{d-1}$ and collapsing it around the operator as in \eqref{u1s2}:
\be
U_{g}(S^{d-1})\sO_b(x) = V^{a}_b(g) \sO_a(x) \label{nonAb} 
\ee
We stress that here the $V$'s are just linear finite-dimensional matrices, and this relation shows how the  quantum operator $U$ acts as a finite dimensional rotation on the charged degrees of freedom. So in the $U(1)$ case e.g. we have that $r$ is a one-dimensional representation and the $V$ is just a phase $e^{i\alpha}$. For $SU(2)$ we might take $r$ to be in the fundamental and then the $V$'s would be exponentials of the Pauli matrices, and so on. 

This may seem unfamiliar, but it is simply a restatement of the regular setup of global symmetries. 
\begin{tcolorbox}

{\bf Exercise} 
\begin{enumerate}
\item Prove, by cutting open the path integral on a time-slice, that \eqref{infu1} is equivalent to the familiar operator equation 
\be
[Q, \sO(\vec{x})] = i \sO(\vec{x})
\ee
where here $Q$ is defined in the conventional way as in \eqref{ordQdef}. 

\item Derive the corresponding operator relation for the finite non-Abelian transformation from \eqref{nonAb}.
\end{enumerate} 

\end{tcolorbox}  

Finally, we should note that essentially the same formalism works for {\it discrete} symmetries. If we have a theory with a {\it discrete} symmetry group $G$, then we do not have a conserved current for it, but we do have a set of topological operators that enact the action of the discrete symmetry as in \eqref{nonAb}. As an example, imagine a real scalar field theory invariant under the $\mathbb{Z}_2$ symmetry $\phi \to -\phi$; then there would exist a topological defect $U$ which has the following action when collapsed on $\phi$:
\be
U_{\mathbb{Z}_2}(S^{d-1})\phi(x) = -\phi(x)
\ee
As there is only a single element in $\mathbb{Z}_2$ and it is its own inverse, we have the simple fusion rule:
\be
U_{\mathbb{Z}_2}(\sM_{d-1}) U_{\mathbb{Z}_2}(\sM_{d-1}) = 1
\ee

The main point of all of this is just to note that symmetries give us topological operators, often called {\bf charge defects}. 

Because the charged objects here are $0$-dimensional local operators, ordinary global symmetries are called  {\bf $0$-form symmetries.}  
\section{Higher form symmetries}

Once we realize that global symmetries essentially define a {\it topological sector} of a theory, we can generalize these in many productive ways. We will first consider {\bf higher-form symmetries}\cite{Gaiotto:2014kfa}, which basically ask the question: if normal symmetries are associated with the conservation of particles, what symmetries are associated with the conservation of {\it extended objects}, such as strings or branes? This is not a question only for string theorists: such objects exist in real life.
 
\subsection{Idea}

We will start with the continuous case. Consider an antisymmetric two-index current $J^{\mu\nu} = J^{[\mu\nu]}$ which is conserved:
\be
\p_{\mu} J^{\mu\nu} = 0 \qquad d \star J = 0
\ee
Just as ordinary 0-form global symmetries count particles, this object is associated with a $U(1)$ symmetry which counts {\it strings}! This is called a {\bf $1$-form global symmetry}. 

\begin{figure}[h]
\begin{center}
\includegraphics[scale=0.5]{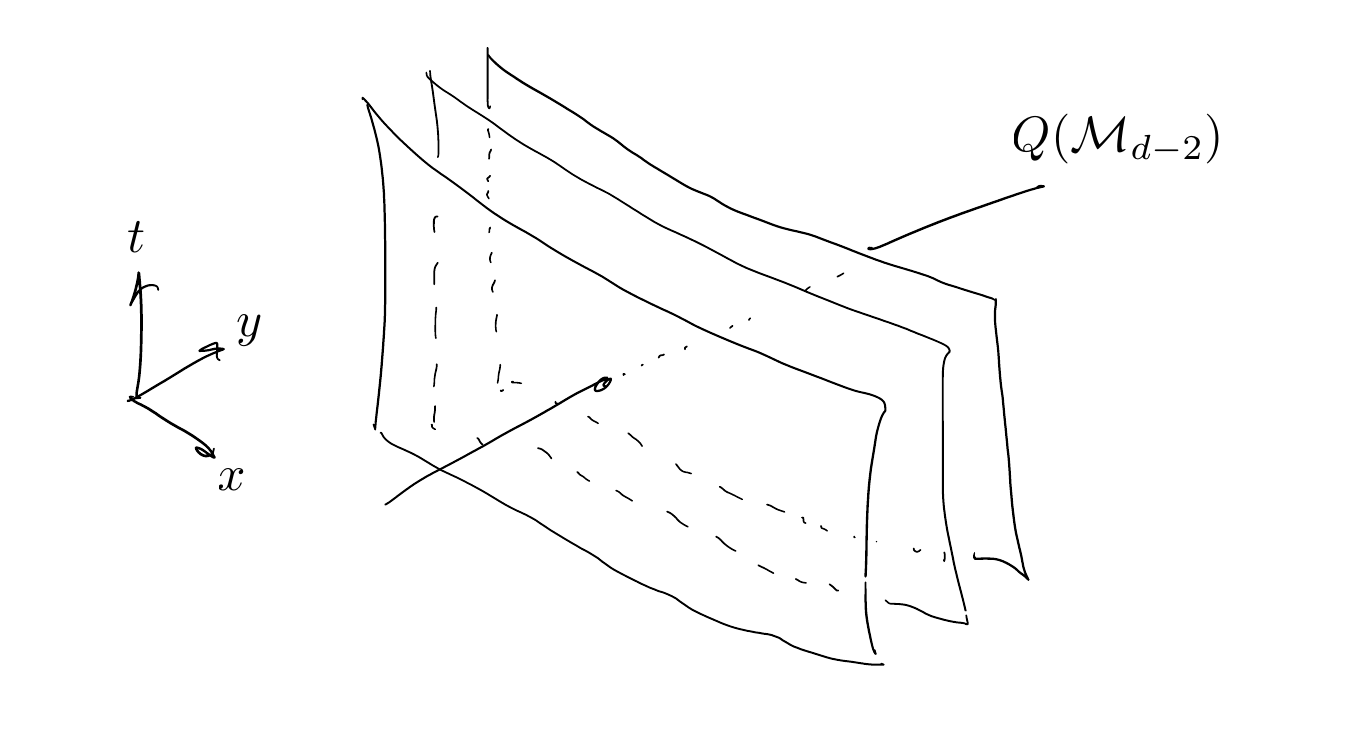}
\end{center}
\caption{To count 2d string worldsheets, one does an integral over a codimension-$2$ surface, not a codimension-$1$ surface as for particle worldlines; compare this to Figure \ref{fig:0formparticles}.}
\label{fig:countingstrings}
\end{figure}

At some rudimentary level the extra index tells us which way the string points. Let's dig into this a little bit; first, following \eqref{Qdef} we can form the following object
\be
Q(\sM_{d-2}) = \int_{\sM_{d-2}} \sqrt{h} n^{\mu\nu} J_{\mu\nu} = \int_{\sM_{d-2}} \star J
\ee
which we integrate over a codimension-$2$ manifold. For the same reasons as above, this is a {\bf topological defect}, which is now codimension-$2$. This is 2-dimensional in 4d. For intuition, imagine a sea of strings -- i.e. 2-dimensional worldsheets without end -- poking through the screen on which you read this, as in Figure \ref{fig:countingstrings}. To count all of them we don't need to integrate over the whole room, we just integrate over the screen -- we can then move the surface $\sM_{d-2}$ upwards in time {\it or} forwards in space and get the same answer, because the strings have no end. 

We have formulated ordinary symmetries in a fancy way above so that we can exploit the fact that much of the formalism carries over very simply. First, what is the charged object? Recall from before that a charged object is the ``thing that creates the thing that you count'', in other words it should be the object who creates a string. 

This is a {\bf line operator $W(C)$} defined on a 1d curve $C \subset \mathbb{R}^d$, which can be imagined as creating a string living along that curve $C$ at an instant in time. If it has charge $q$ then the Ward identity in this case now becomes:
\be
\p_{\mu} J^{\mu\nu}(x)W(C) = iq\int_{C} ds \frac{dX_C^{\nu}}{ds}\delta^{(d)}(x-X_{C}(s)) W(C) \label{ward1form} 
\ee
which is a somewhat clunky way to write that the current is not conserved whenever we evaluate it on the worldline of the extended charged operator. Note that the extra index $\nu$ comes from the geometry of the line operator. 

We can write this much more elegantly if we define the $(d-p)$ form-valued delta function $\delta_{C_p}(x)$ which depends on a $p$-submanifold $C_p$: it is zero whenever $x \notin C_p$ and ``integrates to 1'', which in this case means that
\be
\int_{\mathbb{R}^d} B_p \wedge \delta_{C_p}(x) = \int_{C_p} B_p
\ee
for all test forms $B_p$. The case above is for $p = 1$, and we can write it as
\be
d \star J (x) W(C) = i q \delta_{C}(x) W(C) \label{ward1form} 
\ee
This is just notation to replace the integral in \eqref{ward1form}. 

\begin{figure}[h]
\begin{center}
\includegraphics[scale=0.5]{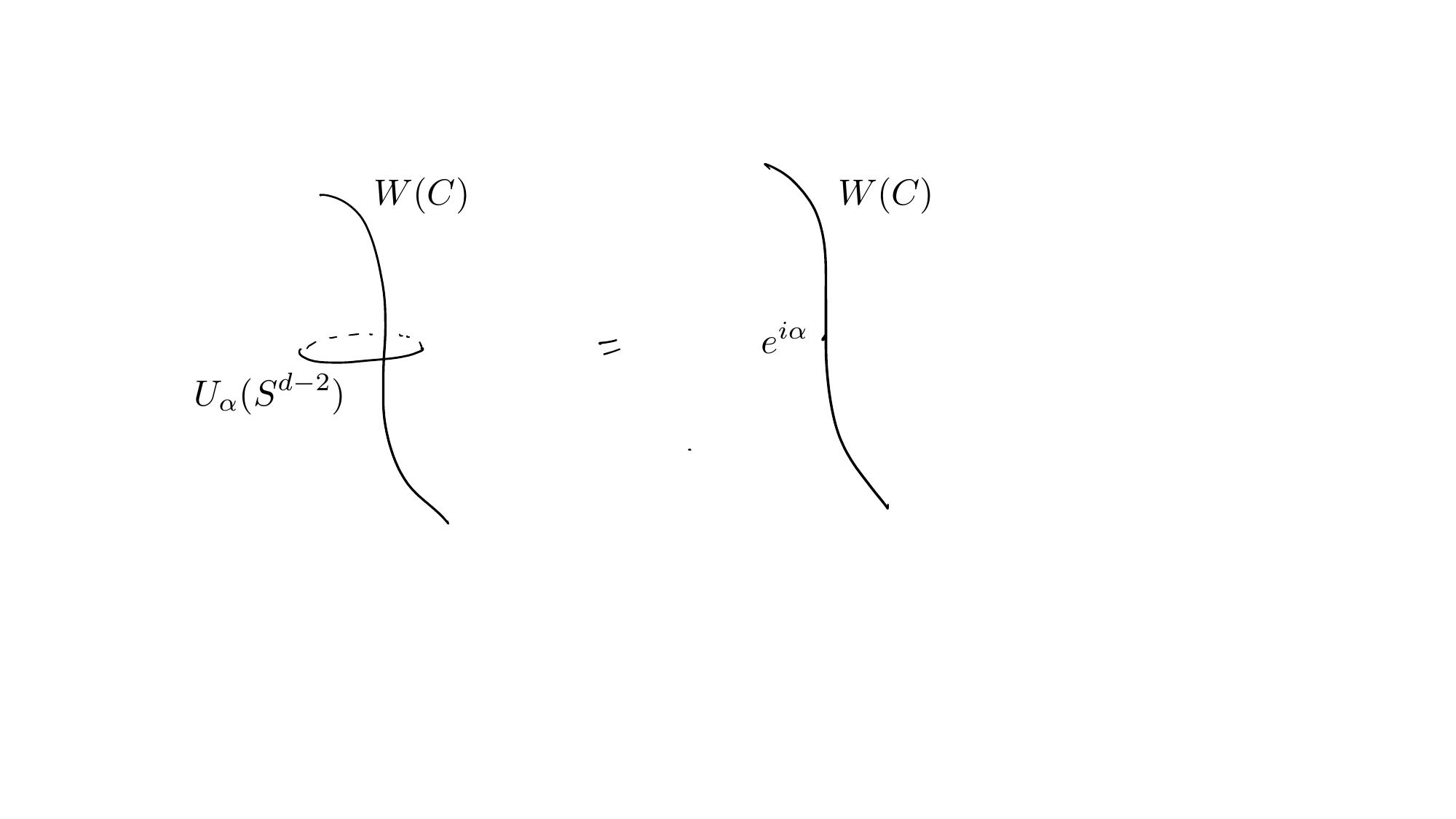}
\end{center}
\caption{Collapsing the 1-form charge defect $U_{\al}(S^{d-2})$ onto a line operator acts as a phase rotation on the line.}
\label{fig:collapseS-1form}
\end{figure}

Now again we can construct the object who generates the finite rotation by an angle $\alpha$:
\be
U_{\al}(\sM_{d-2}) = \exp(i \al Q(\sM_{d-2}))
\ee
We should now ask what happens if we drag this object through a line operator $W(C)$. The idea is the same as in \eqref{movingthrough} -- $W(C)$ will pick up a phase. This is most cleanly represented if we imagine placing the defect on a $S^{d-2}$ that links the $W(C)$ and then collapsing it, as in Figure \ref{fig:collapseS-1form}. 
\be
U_{\al}(S^{d-2}) W(C) = e^{iq\alpha} W(C) \label{collapse1form} 
\ee

\begin{tcolorbox}
{\bf Exercise:} Prove \eqref{collapse1form} from \eqref{ward1form}.
\end{tcolorbox}  
This is the idea of a 1-form symmetry. 

It can clearly be generalized to any $p$-form symmetry, which in the continuous case means that there exists a $p+1$-form symmetry current $J^{\mu_{1}\cdots\mu_{p+1}}$, charge defects $Q(\sM_{d-p-1})$, and charge{\it d} surface operators that exist on submanifolds $C_p$ of dimensionality $p$. A 0-form symmetry counts particles, a 1-form symmetry strings, a 2-form symmetry bedsheet-type objects, and so on.\footnote{In fact the idea can even usefully be generalized in the other direction to a (-1)-form symmetry \cite{Aloni:2024jpb}, at least in the $U(1)$ case, though the notion of ``counting'' things becomes a bit abstract.} 

It cannot however be generalized to the non-Abelian case: to understand this let's consider doing the group composition \eqref{groupcomp} for a $p$-form symmetry. 
\be
U_{g}(\sM_{d-p-1}) U_{g'}(\sM_{d-p-1}) = U_{g g'}(\sM_{d-p-1})
\ee
The ordering on the right hand side of $gg'$ was inherited from the ordering on the left hand side. For codimension $1$ objects (i.e. for $p = 0$) there is a well-defined notion of ordering -- this is the time-ordering that is inherited from the path integral. However for higher codimension objects there isn't -- you can always move one around the other. Thus the ordering is ambiguous and the right hand side of the fusion cannot depend on it, and thus $g$ and $g'$ necessarily commute. 

We can however easily have {\it discrete} higher-form symmetries -- e.g. $\mathbb{Z}_N$ -- which will indeed play an important role below. Just as for 0-form symmetries, we then have a charge operator $U_{g}(\sM_{d-p-1})$ where $g \in \mathbb{Z}_N$. See the recent review \cite{Brennan:2023mmt} for an extensive discussion of discrete higher-form symmetries. 

\subsection{Phases of 1-form symmetry}

Let us now understand the phases of 1-form symmetries. We will do this as we did around \eqref{expdecay} and \eqref{odlro}, i.e. by understanding the behavior of the expectation value of the charged object $\vev{W(C)}$:
\ben
\item {\bf Unbroken phase}: in the unbroken phase, we have that for large closed curves $C$ there is the {\bf area law}:
\be
\vev{W(C)} \sim \exp\le(-T \mbox{Area}[C]\ri)
\ee
where $\mathrm{Area}[C]$ is the area of the minimal surface that fills in the curve $C$, i.e. the minimal surface $S$ such that
\be
\p S = C \ . 
\ee 
This is the higher-form analogue of the exponential decay shown in \eqref{expdecay}. The way to understand this is that the objects that we count -- i.e. the string worldsheets -- are massive with tension $T$; thus the line operator creates a 2d worldsheet which must fill in the curve. 

You will be familiar with the area law as a diagnostic for confinement in non-Abelian gauge theory; we will return to this below. 

\item {\bf Spontaneously broken phase:} here you should imagine that the strings have ``condensed'', i.e. that the tension has dropped to zero. What does this mean? This may seem alarming; however in this case it means that the behavior of the line operator has changed to a {\bf perimeter law}. 
\be
\vev{W(C)} \sim \exp\le(-m L[C] \ri) \label{perimeter} 
\ee
where $L[C]$ is just the length of the curve (or more generally any {\it local} functional of the curve data). The idea here is that just as in \eqref{odlro} the correlation function has factorized {\it as much as possible}; however the line operator can always be redefined by a c-number which is a local functional of $C$, and thus the behavior above is essentially the same as saying that the line operator is a constant which is independent of $C$. The parameter $m$ is thus non-universal. 
\een 

Finally, we could imagine the transition between these phases, where we would have
\begin{enumerate}
\setcounter{enumi}{2}
\item {\bf Phase transition:} if the transition is continuous, we would expect to see a conformal field theory describing it, and here we would have
\be
\vev{W(C)} = \exp\le(-m' L[C] \ri)f[C]
\ee
where $f[C]$ is expected to depend on $C$ in a nontrivial manner, and might capture some interesting universal data determined by the transition. Unfortunately there are not too many continuous transitions involving the spontaneous breaking of a 1-form symmetry; they turn out to usually be first-order\footnote{For a possible explanation see \cite{Iqbal:2021rkn}.}. An important exception is $\mathbb{Z}_2$ gauge theory in 3d, which we will discuss. 
\end{enumerate} 
The analogy with 0-form symmetries should be clear. As we will see, this classification can also be {\it useful} -- this will provide a useful classification of phases, and we can do all of the same sorts of things that we did for 0-form symmetries for their higher-form cousins. 

\begin{figure}[h]
\begin{center}
\includegraphics[scale=0.5]{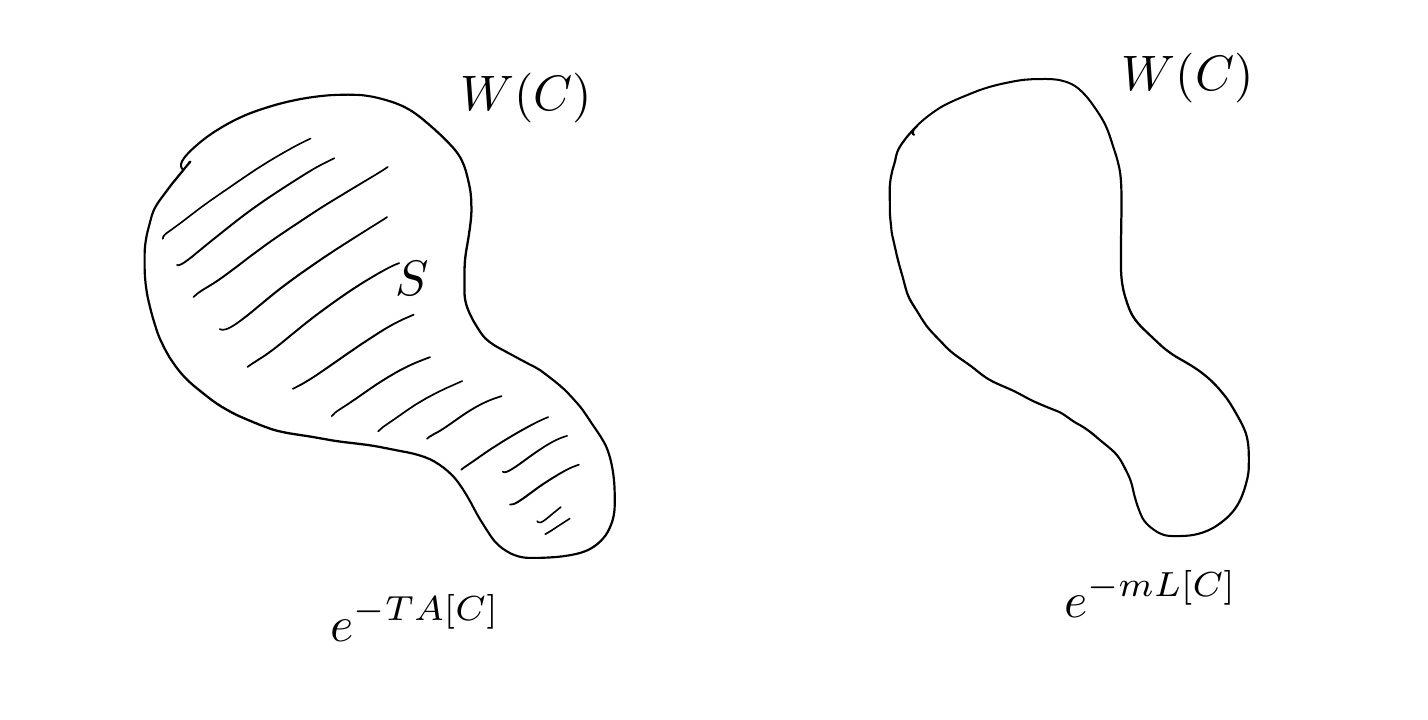}
\end{center}
\caption{Area vs. perimeter laws in the spontaneous breaking of a 1-form symmetry.}
\label{fig:areavperimeter}
\end{figure}

It should be clear how the ideas above generalize to an arbitrary $p > 1$. 

\subsection{Genuine and non-genuine operators}
At this point we have introduced already a few types of extended objects. This is a good time to define some more terminology.

A {\bf topological} operator (which can be either local or an extended defect) is an object which is invariant under small wiggles of its support. An example is the identity operator {\bf 1} and any of the extended symmetry defects defined above. (Indeed, by Schur's lemma, any {\bf local} topological operator is necessarily proportional to the identity). 

A {\bf genuine} operator (which can be either local or extended) is an operator which does not have anything tied to it. Examples are most of the operators you know and love. 

A {\bf non-genuine} operator (which can be either local or extended) is an operator who {\bf does} have something tied to it. Examples include: 
\ben
\item Fields that are charged under gauge symmetries -- e.g. in QED with a charged scalar $\Phi$ necessarily has a Wilson line attached to it, so it is not a genuine local operator: we have to write it as $\Phi(x) e^{i \int_{x} A}$ where the Wilson line defines a curve $C$ that ends on $x$. Note that in this example the Wilson line is usually not topological, which means that you are generally aware that it is there. 
\item A more insidious example of this phenomenon is when the ``something tied'' to the operator is topological. In that case one might miss it and then be tempted to conflate a genuine operator with a non-genuine one, leading to confusion. 
\item Both cases can happen for extended objects: e.g. you can have a 1d line operator with a 2d surface tied to it, where the surface might or might not be topological. 
\een

\subsubsection{Non-genuine local operators: 2d compact boson} 
Let us study an example of a non-genuine local operator. This section requires some familiarity with 2d CFT. Consider the compact boson CFT with action
\be
\frac{1}{2g^2} \int d^2x\;(\p\theta)^2
\ee
where $\th \sim \th + 2\pi$. Let us formulate it on $S^1 \times \mathbb{R}$. In that case there is the well-known state operator correspondence that maps states on the cylinder to local operators on $\mathbb{R}^2$. In particular, the vacuum on the cylinder maps onto the identity on $\mathbb{R}^2$. 

Now we are free to formulate boundary conditions as we like on the $S^1$. It is a reasonable thing to demand that $\th$ wind an integer number of times around the $S^1$, i.e. that
\be
\th(\phi) = \th(\phi+2\pi) + 2\pi n \qquad n \in \mathbb{Z}
\ee
The vacuum under this boundary condition defines a new state $\ket{n}$. We can then map the vacuum with this boundary condition back to the plane by the state-operator correspondence. This defines a family of local operators $V_{n}(x)$ labeled by an integer $n$. These are {\bf genuine} operators. 

One could have formulated it directly on the plane by demanding that in a path integral formulation of the theory, the integral of the 1-form $d\theta$ on a circle wrapping the point $x$ be quantized as $2\pi n$, i.e.
\be
\oint d\theta = 2\pi n
\ee
Note that in this formulation we have an operator $V_n(x)$ who is defined not as a polynomial of fundamental fields in the Lagrangian (which is how we normally define fields) but rather as a boundary condition on fields in the path integral. This is an example of a {\bf disorder operator}, which is still {\bf genuine}. 

Now consider the following wrong-feeling operation: let's consider a boundary condition on the $S^1$ so that $\theta$ has a {\it non}-integer winding:
\be
\th(\phi) = \th(\phi+2\pi) + 2\pi \alpha \qquad \alpha \notin \mathbb{Z}
\ee
We are free to do this; but what does it mean? Local physics on the cylinder proceeds mostly as normal; but if we map back to the plane the putative local operator that we try to define $\tilde{V}_{\alpha}(x)$ will have a branch cut emanating out of it. This branch cut is the location of the seam where the field $\th$ jumps by $2\pi \al$. It is a topological line, as clearly we can move it around mostly as we like. Thus the operator dual to the non-integer winding vacuum has a tail -- i.e. a topological line operator ending on it -- and is a {\bf non-genuine operator}.

\begin{tcolorbox}
{\bf Exercise:} Show that the ``tail'' above is actually the 1d charge defect for the 0-form $U(1)$ global symmetry associated with $\th$ translation. {\bf Hint}: what is the action of the tail on the operator $e^{im\th}$? 
\end{tcolorbox}

\section{Applications}

This has all been rather formal. We will now study a few examples and see what this formalism can do for us. 
\subsection{Free electromagnetism}
Consider familiar free $U(1)$ electromagnetism in 4d with no matter. 
\be
S[A] = \int d^4x \le(\frac{1}{4 e^2} F^2\ri) \qquad F = dA
\ee
The equation of motion is $\p_{\mu} F^{\mu\nu} = 0$. 
The action above does not fully specify the theory, as it describes only the Lie {\it algebra} of the gauge group and not the gauge group itself. 
It will soon be important that we are talking about {\it compact} electromagnetism with gauge group $U(1)$ and not $\mathbb{R}$. This means that gauge transformations are labeled by elements $g = e^{i\Lambda(x)} \in U(1)$, with $A \to g^{-1}(A + d)g$. 

What are the symmetries? Introductory textbooks would now talk about gauge symmetry. 

However, a sophisticated reader might make the argument that gauge ``symmetry'' is a misleading name; gauge symmetry is not really a symmetry at all, but rather a redundant way of describing the dynamics. As we have discussed above, it is entirely possible to reformulate a theory described with a Lagrangian involving gauge fields in terms of a different Lagrangian with no gauge fields at all. So we can (a tad pretentiously) ask: what is the invariant physical information present in this theory? In other words, what are the {\bf global} symmetries? 

This theory has two 1-form global symmetries: an {\bf electric} one $U(1)^{(1)}_e$ and a {\bf magnetic} one $U(1)^{(1)}_m$ with the following currents:
\be
J^{\mu\nu}_e = \frac{1}{e^2} F^{\mu\nu} \qquad  J^{\mu\nu}_m = \frac{1}{4\pi} \ep^{\mu\nu\rho\sig} F_{\rho\sig} \label{Jdef} 
\ee
or in forms
\be
J_e = \frac{1}{e^2} F \qquad J_m = \frac{1}{2\pi} \star F
\ee

The conservation of $J^{\mu\nu}_e$ follows from the equations of motion and is true only on-shell. This captures the idea that {\it electric} field lines are strings that don't end (there is no electrically charged matter). The conservation of $J^{\mu\nu}_m$ follows from the Bianchi identity $\ep^{\mu\nu\rho\sig}\p_{\nu}F_{\rho\sig} = 0$, and captures the idea that {\it magnetic} field lines are strings that don't end (there are no magnetic monopoles). 

What are the charged line operators? 

The line operator charged under $U(1)^{(1)}_e$ is the Wilson line, i.e. the worldline of a test electric charge:
\be
W_{q}(C) = \exp\le(i q \int_{C} A\ri) \label{elWilson} 
\ee 
In the presence of the Wilson line we obtain the expression:
\be
(d \star J_e(x)) W_{q}(C) = i q \delta_{C}(x) W_{q}(C) \label{wardel} 
\ee
These charges are integers $q \in \mathbb{Z}$.  

The action of $U(1)^{(1)}_e$ on the dynamical degrees of freedom can be easily understood in this case -- they correspond to the shift of $A$ by a closed 1-form:
\be
A \to A + \Lambda \qquad d\Lambda = 0 \label{1formshift} 
\ee
which leaves the Lagrangian invariant. Under this the Wilson line operator transforms as a phase
\be
W_{q}(C) \to  e^{iq\int_{C} \Lambda} W_{q}(C) \ .
\ee
How do we couple an external source $b_{\mu\nu}$ to this 1-form symmetry? In the electric case this takes the following form:
\be
S[A;b_e] = \int d^4x \le(-\frac{1}{4 e^2} (dA - b_e)\ri) \label{electricsource} 
\ee
and the 1-form invariance is now enhanced to:
\be
A \to A + \Lambda \qquad b_e \to b_e + d\Lambda \qquad \Lambda\;\;\mbox{arbitrary} \label{maxsource} 
\ee
This is the 1-form analogue of the external coupling to the source that we did back in \eqref{u1extsource}. 

We should note that it is often the case that we cannot identify such a simple action on the degrees of freedom. For example, the line operator charged under $U(1)^{(1)}_m$ is called an {\bf 't Hooft line}, which we will denote by $T_{p}(C)$-- it is the worldline of a {\it test} magnetic monopole. It is an example of a so-called {\it disorder operator} \cite{Kapustin:2005py}, and is defined by demanding that in the path integral over fields, every $S^2$ wrapping the curve $C$ has:
\be
\oint_{S^2(C)} F = 2\pi p \label{magmon} 
\ee
i.e. we have a magnetic monopole. Note that naively we have $F = dA$, which clearly means that the integral on the right-hand side of $F$ must vanish; of course this means that $A$ is not well-defined in the presence of a monopole, and must be defined patch-wise and sewn together on overlapping patches by gauge transformations. The consistency of this sewing together with the integer quantized electric charges in \eqref{elWilson} enforces the Dirac quantization condition,
\be
q p = \mathbb{Z} \label{Dirac} 
\ee
which you will remind yourself of in the exercise below. \eqref{magmon} is equivalent to the following Ward identity for the magnetic current:
\be
(d \star J_m(x)) T_{q}(C) = i q \delta_{C}(x) T_{q}(C) \label{wardmag} 
\ee
For the magnetic symmetry there is no analogue of \eqref{1formshift}. 
\begin{tcolorbox}
{\bf Exercise:} 
\begin{enumerate}
\item Prove the Dirac quantization condition in compact $U(1)$ gauge theory \eqref{Dirac}.
\item Prove the Ward identities \eqref{wardel} and \eqref{wardmag}.  
\end{enumerate} 
\end{tcolorbox} 

\begin{tcolorbox}
{\bf Exercise:} 
\begin{enumerate}
\item Couple an external two-form source $b_m$ to Maxwell theory for the magnetic 1-form symmetry $U(1)^{(1)}_m$, just as we did for the electric 1-form symmetry in \eqref{electricsource}. 
\item Use this to construct a partition function for Maxwell theory $Z[b_e,b_m]$ that is invariant under both 1-form shifts $b_e \to b_e + d\Lam_e, b_m \to b_m + d\Lam_m$. 
\item Actually, this is impossible. This is a manifestation of a {\it mixed 't Hooft anomaly} between the two 1-form symmetries. Develop the theory of this anomaly and explain to what extent each current is conserved. 
\end{enumerate} 
\end{tcolorbox} 

Now that we have understood that we have two 1-form symmmetries $U(1)_{e}^{(1)}$ and $U(1)_{m}^{(1)}$, we should understand how they are realized: i.e. are they spontaneously broken, or unbroken, etc? To do that we  compute the expectation values $\vev{W_q(C)}$ and $\vev{T_p(C)}$. 

The calculation of $\vev{W_{q}(C)}$ for a circular loop is a textbook exercise. Heuristically speaking the Wilson line emits photons but they don't do too much, and the leading contribution comes from a short-distance singularity along the perimeter of the loop, so we find:
\be
\vev{W_{q}(C)} \sim \exp(-\Lambda L[C])
\ee
with $\Lambda$ a non-universal cutoff. This is a perimeter law. $\vev{T_{q}(C)}$ behaves the same way -- one way to understand this is that the action of free electrodynamics is invariant under electric-magnetic duality, which interchanges the Wilson and 't Hooft lines. Thus in the normal phase of ``free EM'' we see that {\it both higher form symmetries are spontaneously broken}. 

This is an important fact -- the way to think about it that the vacuum is a sea of ``condensed'' strings, where the condensed objects are magnetic and electric field lines. Long-wavelength wiggles in this sea of condensed strings are what we call photons. Indeed this allows us to ask the question: {\bf why is the 4d photon massless?} 

When first learning QFT we often hear some words about why the photon is massless due to gauge invariance, which sounds pretty good until the next chapter where we learn about the Higgs mechanism, which writes down a perfectly gauge-invariant action that gives the photon a mass. Thus the conventional way of thinking about this is somewhat unfulfilling. 

The modern way of thinking about this is that the photon is the Goldstone mode of a spontaneously broken higher-form {\it global} symmetry\footnote{See \cite{Kovner:1992pu} for a very early discussion of this idea.}. We should thus prove Goldstone's theorem for higher-form symmetries.

\subsection{Interlude: Goldstone's theorem}
Before proving Goldstone's theorem for higher-form symmetries, we will first revisit the proof for ordinary 0-form symmetries. The canonical non-perturbative proof of Goldstone's theorem involves Hamiltonian methods \cite{Goldstone:1962es}; here we will discuss a Euclidean path-integral proof, following \cite{Hofman:2018lfz}. 
\subsubsection{Goldstone's theorem for 0-form symmetries}
\begin{figure}[h]
\begin{center}
\includegraphics[scale=0.5]{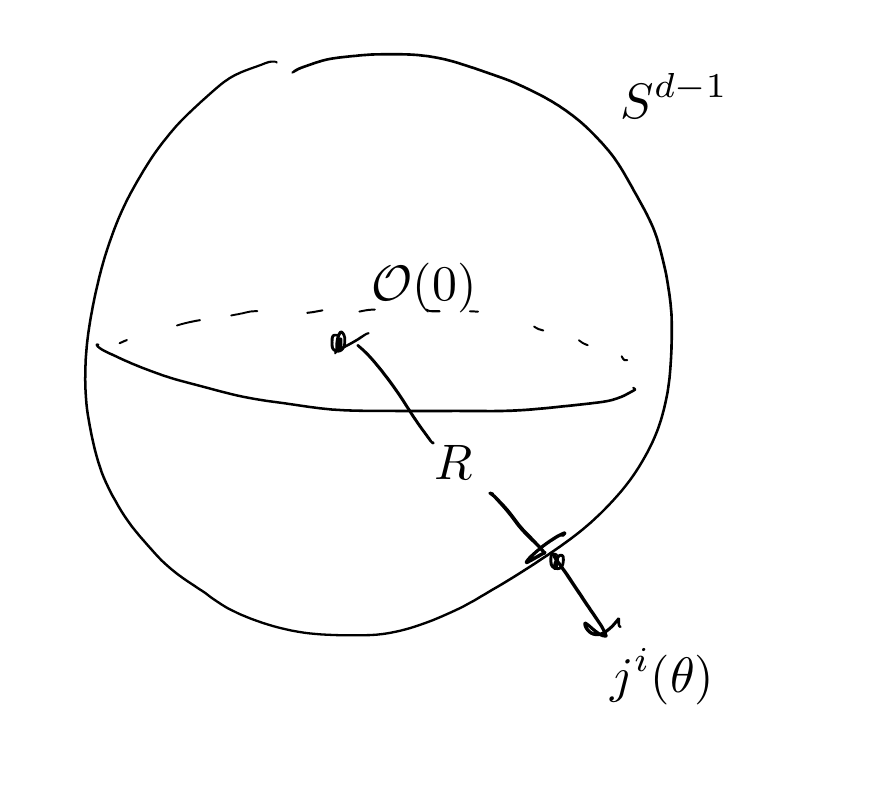}
\end{center}
\caption{Geometry used for proof of 0-form Goldstone theorem.}
\label{fig:goldstone0form}
\end{figure}

 Let us begin by considering the Ward identity for a charged operator $\sO(x)$ at the origin. 
\be
\p_{\mu} j^{\mu}(x) \sO(0)  = iq \delta^{(d)}(x) \sO(0) 
\ee
Let us now integrate $x$ on both sides over a ball $B^d$ which is the interior of a $S^{d-1}$ of radius $R$ centered at the origin, as in Figure \ref{fig:goldstone0form}. 

We find using the divergence theorem that
\be
\oint_{S^{d-1}(R)} d^{d-1}\th \;n_i j^i(\theta) \sO(0) = iq\int_{B^d(R)} d^dx \delta^{(d)}(x) \sO(0)
\ee
where $\th$ describes the angles parametrizing the $S^{d-1}$. Now the right hand side integral is trivial and picks up the delta function at the origin, giving us
\be
\oint_{S^{d-1}(R)} d^{d-1}\th \;n_i j^i(\theta) \sO(0) = iq \sO(0)
\ee
So far this has been true in any theory, independent of the phase. Now take the expectation value of both sides. We find
\be
\bigg\langle\oint_{S^{d-1}(R)} d^{d-1}\th \;n_i j^i(\theta) \sO(0)\bigg\rangle = iq \vev{\sO} \ . 
\ee
There are now two possibilities:
\ben
\item The symmetry is {\bf unbroken}, so $\vev{\sO} = 0$. The left-hand side also vanishes. In this case we can say nothing else. 
\item The symmetry is {\bf spontaneously broken}, so $\vev{\sO} \neq 0$. In this case the left-hand side is both nonzero and independent of the radius $R$. Thus by spherical symmetry we have that the correlation function between the the operator at the origin and the current at a distance $R$ from it satisfies:
\be
\vev{j^i(\theta)\sO(0)} \sim \frac{iq \vev{\sO}n^i}{R^{d-1}}
\ee
(Amusingly, this is exactly how we derive the classic $r^{-2}$ potential using Gauss's law in elementary EM, which is not a coincidence). Thus there is a power-law in distance correlation in the theory, and there must be at least one gapless mode to generate this power-law correlation. This gapless mode is the Goldstone mode. 
\een

This is the fastest proof of the ordinary Goldstone theorem, though to our knowledge it was first written down in \cite{Hofman:2018lfz}. Though it is formulated in Euclidean language, it is conceptually equivalent to the ordinary Hamiltonian proof; in particular if you cut open the path integral then you find the commutators that one normally uses. 

\subsubsection{Goldstone's theorem for higher form symmetries}

\begin{figure}[h]
\begin{center}
\includegraphics[scale=0.5]{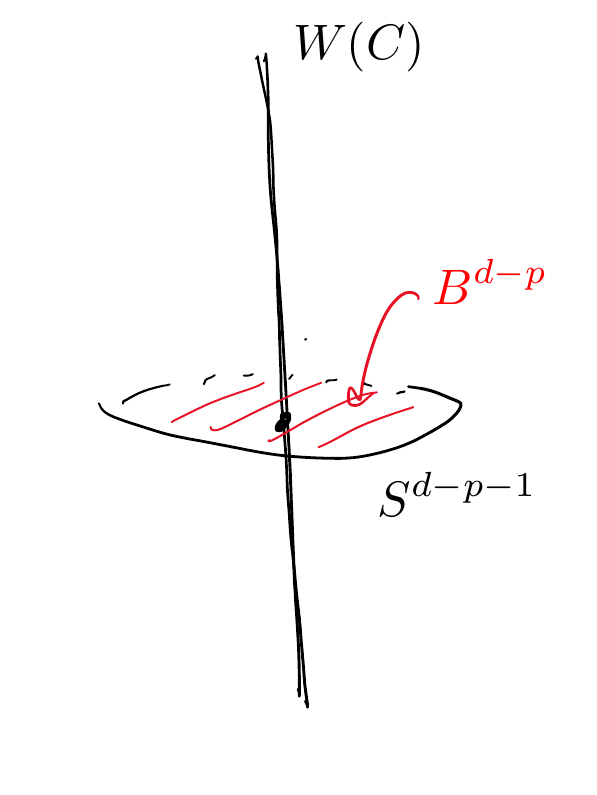}
\end{center}
\caption{Geometry used for proof of p-form Goldstone theorem.}
\label{fig:goldstone1form}
\end{figure}

We now present the proof for higher form symmetries. It is really exactly the same as above.\footnote{For a proof using Hamiltonian methods, see \cite{Lake:2018dqm}.} We begin by considering the Ward identity for a $p$-form symmetry:
\be
d \star J (x) W(C) = i q \delta_{C}(x) W(C)
\ee
Let us now assume that we are in a spontaneously broken phase, as that is the case of interest. In that case $W(C)$ obeys a perimeter law as in \eqref{perimeter}
\be
\vev{W(C)} \sim \exp(-m \mbox{Perimeter}[C])
\ee 
where here ``perimeter'' denotes the $p$-volume of the submanifold $C$. Note that as this is a local functional of the geometric data characterizing $C$, we can define a new charged operator $\overline{W}(C)$ that strips this off, 
\be
\overline{W}(C) \equiv \exp(+m \mbox{Perimeter}[C])W(C)
\ee
$\overline{W}(C)$ satisfies the same Ward identity as $W(C)$, i.e.
\be
d \star J (x) \overline{W}(C) = i q \delta_{C}(x) \overline{W}(C) \label{wardstripped} 
\ee
 Now just as above, we take $C$ to be an infinite flat $p$ plane and consider a $(d-p)$ dimensional ball $B^{d-p}$ that intersects $C$ at a single point. This ball has a boundary $S^{d-p-1}$ and the perpendicular distance from the intersection point to the $S^{d-p-1}$ is $R$, as in Figure \ref{fig:goldstone1form}. 
 
Now we integrate $x$ on both sides of \eqref{wardstripped} over the $B^{d-p}$. On the left hand side we have
\be
\int_{B^{d-p}} d \star J (x) \overline{W}(C) = \oint_{S^{d-p}(R)} \star J(\theta) \overline{W}(C) \ . 
\ee
where $\theta$ is a point on the $S^{d-p}$. On the right hand side the delta function contributes from a single point, so we have
\be
\oint_{S^{d-p}(R)} \star J(\theta) \overline{W}(C) = i q \overline{W}(C) 
\ee
Finally we take the expectation value:
\be
\big \langle \oint_{S^{d-p}(R)} \star J(\theta) \overline{W}(C) \big \rangle = i q \vev{\overline{W}(C)}
\ee
By construction $\vev{\overline{W}(C)}$ is nonzero, and thus the left hand side is both nonzero and independent of the radius $R$, which means that the correlation function must behave as
\be
\langle\star J(\theta) \overline{W}(C)\rangle \sim \frac{i q}{R^{d-p-1}} \label{brokenWard} 
\ee
i.e. there is a power-law correlation between the current at a distance $R$ from the surface operator. This power law correlation means that there is a single gapless mode in the spectrum, which is the Goldstone mode.

One can now ask what the low-energy theory is that describes this Goldstone mode, i.e. what is the higher-form analogue of \eqref{0formGoldstone}? It will not surprise you that this is given by a $p$-form gauge theory, i.e.
\be
S = \int d^{d}x \left(v^2 (dB_{p})^2\right)  \label{lowenergac}
\ee
where the extended charge operator is realized in terms of this degree of freedom by
\be
W(C) = \exp\left(iq \int_{C} B_p\right) \label{lowenergsurf} 
\ee
In the case $p = 1$ this is exactly the structure of ordinary electromagnetism! If you run the argument with $J_e$ then the low-lying Goldstone mode that you get is the ordinary photon $B_{p=1} = A$ and \eqref{lowenergsurf} is exactly the ordinary Wilson line. If you run the argument with $J_{m}$ then the Goldstone mode turns out to be the magnetic photon $B_{p=1} = \tilde{A}$, which is related to the ordinary one by electric-magnetic duality; then \eqref{lowenergsurf} turns out to represent the 't Hooft line instead. 

\begin{tcolorbox} 
{\bf Exercise:} Prove in the first non-trivial case $p=1$ that the low-energy effective action \eqref{lowenergac} saturates the spontaneously broken Ward identity \eqref{brokenWard}. 
\end{tcolorbox} 

This is maybe the first case where we see that the structures from ordinary symmetries really do carry over to the higher-form case; we really obtain nontrivial dynamical information, i.e. from the behavior of a certain line operator we can now state that there exists a gapless mode in the spectrum.\footnote{See \cite{Hidaka:2020ucc} for a discussion on how the counting of higher-form Goldstone modes can be modified when they mix with other fields that break spacetime symmetries.} 

\subsection{Electrodynamics coupled to matter} \label{sec:matter} 
Now that we understand that the free photon is the Goldstone mode of a spontaneously broken 1-form symmetry, we should think about what happens if we couple in electric matter. Let's start with a scalar of integer charge $q$, i.e. the action is now: 
\be
S[\phi, A] = \int d^4x\le(\frac{1}{4e^2} F^2 + (D\phi)^{\dagger} D\phi + V(\phi^{\dagger}\phi) \ri)
\ee
with the usual gauge-covariant derivative $D_{\mu}\phi = \p_{\mu} \phi - i q A_{\mu}\phi$ and where we take
\be
V(\phi^{\dagger}\phi) = m^2 \phi^{\dagger}\phi + \frac{\lambda}{4} (\phi^{\dagger}\phi)^2 + \cdots
\ee
Let us now ask what has happened to the symmetry currents we identified before. 
We begin by studying the case $q = 1$. The equations of motion now say that
\be
\p_{\mu} J^{\mu\nu}_{e} = iq \le(\phi^{\dagger} D^{\mu} \phi - \phi D^{\mu} \phi^{\dagger}\ri) = j^{\mu} _{\mathrm{el}} \label{enoncons} 
\ee
i.e. the electric 1-form symmetry $U(1)^{(1)}_e$ has been {\bf explicitly} broken as the corresponding current is no longer conserved. Physically, the point is that electric field lines can now end on free electric charges. 

There is another way to see that $U(1)^{(1)}_e$ is now broken. The charged Wilson line operators can now end on scalar field insertions as $\phi(x) \exp\le(i\int_{x}^{\infty} ds A\ri)$. The line operator then cannot have a nontrivial linking with any topological charge operator as we can just pull the charge operator off the end and take it away. 

The magnetic 1-form symmetry $U(1)^{(1)}_m$, whose current is:
\be
J^{\mu\nu}_{m} = \ha \ep^{\mu\nu\rho\sig} F_{\rho\sig}
\ee
remains, as there are no dynamical magnetic monopoles in this theory; of course it may still be spontaneously broken. To understand this, we should separate the two infrared phases of this gauge theory:
\ben
\item {\bf Coulomb phase:} this is when the mass in the potential $m^2 > 0$, and so the scalar $\phi$ is uncondensed. In that case if we go to very long distances we can essentially ignore the $\phi$ field completely, as the right hand side of \eqref{enoncons} is energetically suppressed. We are then back in the situation of free EM. Note that the electric 1-form symmetry looks like it has come back -- it is {\bf emergent} at long distances, and then spontaneously broken. This sort of situation -- i.e. the emergence of a symmetry at long distances in the absence of fine-tuning -- is perhaps somewhat rare for 0-form symmetries but is generic for 1-form symmetries. The magnetic 1-form symmetry is also spontaneously broken. 

\item {\bf Higgs phase:} here we imagine adjusting the potential by taking $m^2 < 0$ so that the gauge-charged scalar condenses: this is then the Higgs or superconducting phase. 

Then it's convenient to write $\phi(x) = v(x) e^{i\theta(x)}$ and the low-energy physics is described by:
\be
S = \int d^4x \left(\frac{1}{4 e^2} F^2 + v_0^2 (d\theta - A)^2 + (d v)^2 + m_{H}^2 (v - v_0)^2 + \cdots \right) \label{higgsac}
\ee
It is well-known that in this Higgs phase the spectrum is completely gapped. What has happened to our 1-form symmetries? The electric one is completely broken; in this phase you definitely cannot ignore the right hand side of \eqref{enoncons}. 

To understand how the magnetic symmetry is realized, we should compute $\vev{T(C)}$ in this phase. We are thus asking: what happens if we place a probe magnetic monopole into a Higgs phase? In the Coulomb phase the magnetic field lines can spread out. In the Higgs phase, however, they cannot -- the very point of the Higgs phase is that magnetic fields are expelled. If you force flux in nevertheless, the magnetic field lines are forced into a tight little bundle called the Abrikosov flux tube. 

Let us briefly describe the field configuration of the flux tube: we are searching for a solution to the equations of motion of \eqref{higgsac} that supports nonzero magnetic flux. It's helpful to go to cylindrical coordinates
\be
ds^2 = d\tau^2 + dz^2 + dr^2 + r^2 d\phi^2
\ee
and imagine that we are trying to put a nonzero magnetic field in $F_{r\theta}$. We then find the following configuration. The phase of the scalar field will wind once around:
\be
\theta(r \to \infty) = \phi
\ee
This is terribly singular at the origin, and thus the scalar field will thus necessarily vanish there and interpolate back to its background value at infinity. 
\be
v(r \to 0) \approx 0 \qquad v(r \to \infty) = v_0 \qquad \theta(r \to \infty) = \phi
\ee
This interpolation will take place over a scale set by the Higgs mass $m_{H}^{-1}$. Finally, because the action contains a kinetic term like $(\p_{\phi} \theta - A_{\phi})^2$, at infinity $A_{\phi}(r)$ must track $\theta$: 
\be
\qquad A_{\phi}(r \to \infty) = 1
\ee
Thus the amount of magnetic flux stored in the tube is 
\be
\int_{\mathbb{R}^2} F = \oint_{S^1(\infty)} d\phi A_{\phi} = 2\pi
\ee
and there is thus a single unit of flux. 

Note that this quantization can be traced back to the fact that the field $\theta$ is winding; by creating an integer winding $\th = \mathbb{Z}\phi$ one can create configurations supporting an integer number of units of flux, but there is no way within this phase to create (e.g.) half a unit. 

This configuration is a 2d worldsheet that fills all of $(\tau, z)$ and is localized at $r = 0$; it has an tension $t_p$ per unit worldsheet area which can be calculated numerically and is nonzero. 

Finally, let us return to our basic question: what happens if we insert a 't Hooft line into this theory? The 't Hooft line is a magnetic monopole, which is the endpoint of an Abrikosov vortex -- in other words it will create magnetic flux which is then forced into the 2d worldsheet described above, which will result in an {\it area law}. We thus have
\be
\vev{T(C)} \sim \exp(-t_{p} \mbox{Area}[C]) 
\ee
and the 1-form symmetry in the Higgs phase is {\bf unbroken}. 
\een  
This may all seem backwards -- isn't the gauge symmetry spontaneously broken in the Higgs phase, or something like that? The issue here is really that the previous statement -- though illuminating in the limit of weak gauge coupling $e \to 0$ -- does not really make sense nonperturbatively, because gauge symmetry is not a symmetry. (For example: what would the order parameter be for its breaking? There is nothing observable that is charged under the gauge symmetry, and thus the very idea of an order parameter does not make sense). 

We can summarize the situation in the following table:
\begin{table}[h]
\centering
\begin{tabular}{c|c|c}
& $U(1)^{(1)}_e$ & $U(1)^{(1)}_m$ \\
\hline
Free electrodynamics & Spontaneously broken & Spontaneously broken \\
Massive electrically charged matter & Explicitly broken (emergent in IR) & Spontaneously broken \\ 
Higgs phase & Explicitly broken & Unbroken \\
Massive magnetically charged matter & Spontaneously broken & Explicitly broken (emergent in IR) \\
Confined phase & Unbroken & Explicitly broken 
\end{tabular}
\caption{Realization of 1-form symmetries in different phases of 4d $U(1)$ gauge theory}. 
\label{EMsym} 
\end{table}

To understand the last lines, note that in 4d $U(1)$ electrodynamics we can imagine also coupling in dynamical magnetic monopoles. If they condense then we would end up with a {\it confined} phase, where it is the Wilson line who has an area law. 

Finally, we can consider a non-trivial minimal charge $q > 1 \in \mathbb{Z}$. In this the electric 1-form symmetry is explicitly broken down to $\mathbb{Z}_{q}^{(1)}$; the idea is that electric flux tubes with flux less than the minimal charge can't be screened and so their number is conserved mod $q$. 

%\subsection{Formal interlude: there is no such thing as strongly coupled EM in 4d}
\subsection{Gauge theory: $SU(N)$}
We now turn to non-Abelian gauge theory. 

\subsubsection{SU(N) gauge theory in the continuum} 

We will begin by thinking about pure $SU(N)$ gauge theory. The continuum action depends only on the Lie algebra $\mathfrak{su}(N)$
\be
S[A] = \frac{1}{2g^2}\int d^4x  \Tr(F_{\mu\nu} F^{\mu\nu}) \label{contsun} 
\ee
where the gauge potentials $A_{\mu}$ and field strength $F_{\mu\nu}$ are both valued in $\mathfrak{su}(N)$. As it turns out, the structure of allowed line operators depends intricately not only on the Lie algebra, but also on the global form of the gauge group -- e.g. $SU(N)$ and $SU(N)/\mathbb{Z}_N$ are different groups with the same algebra, and result in corresponding Yang-Mills theories with the same action that are nevertheless subtly different \cite{Aharony:2013hda}\footnote{In fact this sort of ambiguity is present even in the very familiar Standard Model \cite{Tong:2017oea}.}.

It turns out that any pure Yang-Mills theory with gauge group $G$ has (at least) a 1-form symmetry given by the {\it center} of the gauge group. Recall that the center $Z(G)$ of a group $G$ is the set of elements that commute with all elements of $G$, i.e.
\be
Z(G) = \{ g \in G \mid gg' = g'g\;\;\mbox{for every } g' \in G \}
\ee
For concreteness, in these notes we will only discuss pure $SU(N)$ gauge theory. For $SU(N)$, we have that $Z(G) = \mathbb{Z}_N$, written as
\be
\begin{pmatrix}
e^{\frac{2\pi i k}{N}} & & & \\
& e^{\frac{2\pi i k}{N}} & & \\
& & \cdots &  \\
& & & e^{\frac{2\pi i k}{N}} \\
\end{pmatrix} \qquad k \in \{0, \cdots (N-1) \}
\ee
which clearly commutes with everything and has unit determinant. 

Thus the statement is that pure $SU(N)$ gauge theory has a $\mathbb{Z}_N$ 1-form symmetry. 

To understand this 1-form symmetry, we have to do two things: understand the charged line operators, and understand the existence of the topological 2-dimensional surface operator. 

The line operators are -- unsurprisingly -- Wilson lines:
\be
W_{r}(C) = \mathrm{\Tr}_{r} P\le[\exp\le(\oint_{C} A\ri)\ri] 
\ee
Here $r$ denotes the representation. We will see in a second that the charged line operator that we want comes from the fundamental representation $r = f$. 

The 2d topological surface operator is called a {\bf Gukov-Witten surface operator} $G_{g}(\sM_2)$ \cite{Gukov:2006jk,Gukov:2008sn} and in this case it is labeled by an element of $g \in \mathbb{Z}_N$. We will first study a perhaps slightly heuristic explanation of its existence in the continuum: a more convincing argument will require us to study a UV regularization on the lattice. 

The heuristic idea is the following: $G_{g}(\sM_2)$ is defined by saying that when it is inserted into the path integral, the boundary conditions on the gauge field are modified so that holonomy of the gauge field on an infinitesimal loop $C$ that winds around the codimension-2 surface $\sM_2$ is in the same conjugacy class as a center element $g$, i.e.
\be
P \exp\le(\oint_{C} A\ri) = g
\ee
(One can also define these by the conjugacy classes of other group elements that are not in the center; in this case they are however generically not topological objects). 

Physically, these objects can be thought of as inserting vortices which induce a kind of holonomy for the gauge charged objects that move around them. This holonomy acts as a phase on the fundamental Wilson line, i.e. if $g = e^{\frac{2\pi i k}{N}}$ then we have \cite{Heidenreich:2021xpr}:
\be
G_{g}(S^2) W_{f}(C) = \frac{\Tr_{f}(g)}{\Tr_{f}({\mathbf 1})} W_{f}(C) = e^{\frac{2\pi i k}{N}}W_{f}(C) \label{wilsonlink} 
\ee
and thus we see that it generates the desired linking equation \eqref{collapse1form} when a Wilson line is inserted into the path integral. 

(To understand the second equality, note that the holonomy constraint above simply always contributes an extra constant factor of $g$ inside the path-ordered exponential; as $g$ commutes which everything it is proportional to the identity inside the trace and contributes at most a phase, and we can extract that phase by writing it as the ratio of traces). 

However we are looking at pure Yang-Mills theory, where there are no dynamical degrees of freedom transforming in the fundamental. Do any {\it dynamical} degrees of freedom see this surface operator? The dynamical degrees of freedom are the gluons etc. which are in the adjoint. The holonomy picked up by an adjoint-valued degree of freedom as it traverses the Gukov-Witten operator is given by the Wilson line in the adjoint representation:
\be
G_{g}(S^2) W_{\mathrm{Adj}}(C) = \frac{\Tr_{\mathrm{Adj}}(g)}{\Tr_{\mathrm{Adj}}({\mathbf 1})} W_{\mathrm{Adj}}(C) = W_{\mathrm{Adj}}(C)
\ee 
However, as the adjoint representation acts as conjugation $M \to U^{\dagger} M U$, the center doesn't act at all on the adjoint representation, and the trace of any central element in the adjoint is the same as the trace of the identity, $\Tr_{\mathrm{Adj}}(g) = \Tr_{\mathrm{Adj}}(\mathbf{1})$. Thus none of the gluons can see the Gukov-Witten surface operator, and it is topological. 

In the above discussion one may worry that some subtleties are lost in the continuum; we will thus now present the same argument with a lattice description of the gauge theory. This has the benefit of making everything extremely transparent, and we will also see how we can identify discrete symmetries in a lattice model. 

\subsubsection{SU(N) gauge theory on the lattice}
In this section we will formulate pure $SU(N)$ gauge theory on the lattice. This is a standard subject, and our discussion follows that of \cite{tong2018gauge}. Consider a hypercubic lattice $\mathbb{Z}^4$. The basic degree of freedom in lattice gauge theory is a group element $U_{l}\in SU(N)$ defined on each of the directed {\it links} $l$ of the lattice. We will sometimes use a notation $l = \vev{ij}$ to denote a link from site $i$ to site $j$. Reversing the orientation of the link corresponds to taking the inverse of the group element, i.e.
\be
U_{\vev{ij}} = U^{\dagger}_{\vev{ji}} \label{contA} 
\ee
Gauge transformations $\Lambda_i$ are $SU(N)$ valued functions defined on the sites of the lattice, and they act on the link variables as
\be
U_{\vev{ij}} \to \Lambda_{i} U_{\vev{ij}} \Lambda_{j}^{\dagger} \label{latticegauge} 
\ee
Morally speaking, to make the connection to the continuum gauge potential $A$, we should view $U$ as an infinitesimal open Wilson line from $i$ to $j$:
\be
U_{\vev{ij}} = \exp\le(i \int_{i}^{j} A\ri) \ . 
\ee
The analogue of the 2-form field strength is defined on a plaquette, and is constructed by taking the oriented trace of the four links on the plaquette:
\be
W_{\Box} = \mathrm{Tr}_{f} \prod_{l \in \Box} U_{l} \label{wboxdef} 
\ee
Unpacking this for a plaquette in the $(x,y)$ plane with bottom left corner at $(x,y)$ we have:
\be
W_{\Box} = \mathrm{Tr}_{f} \left(U_{\vev{(x,y),(x+1,y)}} U_{\vev{(x+1,y),(x+1,y+1)}} U_{\vev{(x+1,y+1),(x,y+1)}} U_{\vev{{(x,y+1),(x,y)}}}\ri)
\ee
Note that this is invariant under \eqref{latticegauge} because it forms a tiny closed loop and all of the $\Lambda_i$'s cancel. 

We are now ready to form an action. The {\bf Wilson action} for pure Yang-Mills is the following sum over plaquettes:
\be
S[U] = -\frac{1}{g^2} \sum_{\Box}\le(W_{\Box} + W_{\Box}^{\dagger}\ri)
\ee
Here $g^2$ is the Yang-Mills coupling.\footnote{This discussion may make it seem that a naive and generic discretization preserves all the global symmetries of the continuum action. This is not true -- the ``magnetic'' symmetry can be broken by the existence of dynamical defects at the lattice scale \cite{Polchinski:2003bq} -- and this is at the heart of phenomena such as Polyakov's description of confinement in (2+1)d $U(1)$ gauge theory on the latttice \cite{Polyakov:1975rs} and the Berezinskii-Kosterlitz-Thouless transition in 2d \cite{berezinskii1971destruction,kosterlitz2018ordering}. Recent work has shown how these symmetries can be preserved on the lattice, see. e.g. \cite{Sulejmanpasic:2019ytl}.}.  

\begin{tcolorbox}
{\bf Exercise:} By assuming that each link has length $a$ and appropriately Taylor expanding the continuum field $A_{\mu}(x)$ defined in \eqref{contA}, show that to leading order in derivatives
\be
W_{\Box} = -\frac{a^4}{2} \mathrm{Tr}_{f} F_{\mu\nu} F_{\mu\nu} + \cdots
\ee
where $F$ is the usual non-Abelian field strength and there is no sum over indices on the left-hand side (rather the $\mu\nu$ indicate which plaquette we are discussing). Thus show that the lattice Wilson action reduces to the regular continuum action \eqref{contsun} to leading order in derivatives. 
\end{tcolorbox} 

The partition function is then given by integrating over all the group elements on each link:
\be
Z = \int [dU_l] \exp(-S[U]) \label{latticepart}
\ee
There exists a well-defined invariant integration measure on group elements called the {\it Haar measure}, which basically does exactly what you expect it to do and is invariant under the group action. The partition function is a discretely infinite number of integrals over $U$ and can be (fruitfully) done on a computer. 

Now, let us understand the global symmetries of the theory defined by \eqref{latticepart}. The fundamental Wilson line is defined in the obvious way, by taking products of $U$ along a closed path $C$:
\be
W_{r}(C) = \mathrm{Tr}_r\le(\prod_{l \in C} U_{l}\ri) \label{latticeWilson} 
\ee
Let us now turn to the Gukov-Witten surface operator.\footnote{I am thankful to T. Sulejmanpasic for discussions on the content of this section.}  

\begin{figure}[h]
\begin{center}
\includegraphics[scale=0.5]{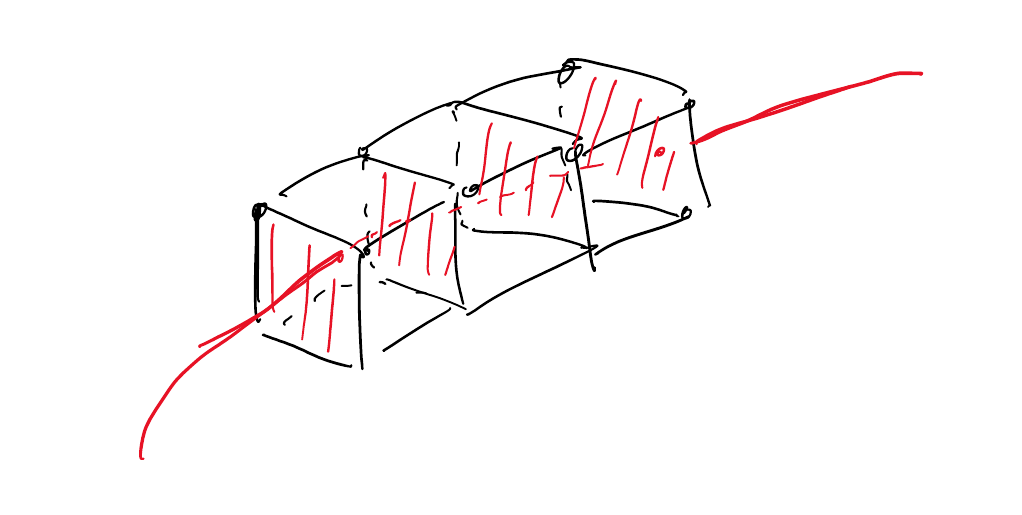}
\end{center}
\caption{A codimension-$2$ surface on the dual lattice intersects plaquettes of the original lattice, shown here in a 3d example; red line represents codimension-$2$ surface, and shaded red plaquettes are intersected by it.}
\label{fig:codim-2-example}
\end{figure}

To begin, note that the Wilson action that we defined above has a natural ``slot'' to which we can couple a $\mathbb{Z}_N$-valued 2-form source. Consider defining a $\mathbb{Z}_N$ valued function on the plaquettes of the lattice by $b_{\Box} \in \mathbb{Z}_N$. Then we can construct a deformed version of $W_{\Box}$ as
\be
W_{\Box}(b_{\Box}) = \mathrm{Tr}_{f} \le(b_{\Box} \prod_{l \in \Box} U_{l} \ri) \ . 
\ee
i.e. we are twisting the product of $U$'s by the appropriate $b$ on the plaquette. It doesn't matter where we place the $b$ in the string of $4$ $U$'s, as it commutes with everything. The action and partition function may be defined in the presence of this source. 
\be
S[U;b] = -\frac{1}{g^2} \sum_{\Box}\le(W_{\Box}(b_{\Box}) + W_{\Box}^{\dagger}(b_{\Box})\ri) \qquad Z[b] = \int [dU_l] \exp(-S[U;b]) \ee
Now to define the Gukov-Witten operator $G_{g}(\sM_{2})$: let $\sM_2$ be a 2d surface on the {\it dual-lattice}\footnote{The dual lattice has a precise definition, but for the cubic lattice its the set of all points that are ``halfway between'' the points of the lattice, i.e. $(\ha, \ha, \ha, \ha)$ + (integer offsets)}. A codimension-2 surface naturally intersects a set of plaquettes on the original lattice (as shown in a 3d example in Figure \ref{fig:codim-2-example}).

\begin{figure}[h]
\begin{center}
\includegraphics[scale=0.5]{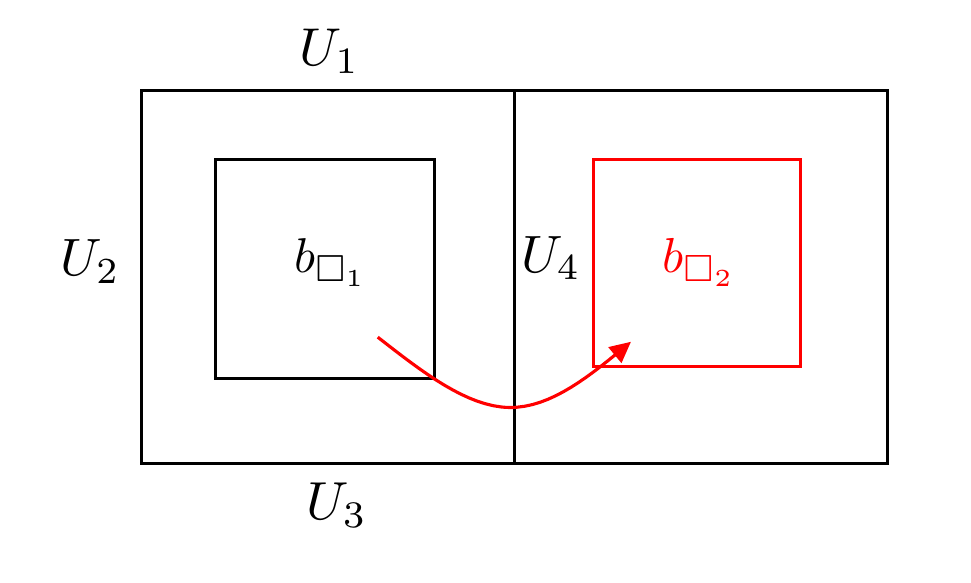}
\end{center}
\caption{Moving the Gukov-Witten surface operator around by a change of variables in the lattice path integral.}
\label{fig:plaquetteBfield}
\end{figure}

For each plaquette that intersects $\sM_2$, set $b_{\Box} = g$. (For all other plaquettes, set $b_{\Box} = \mathbf{1}$). The resulting choice of $b_{\sM_2}$ can be thought of as defining a new partition function which depends on a choice of surface corresponding to the insertion of the operator\footnote{For an interesting study in 2d of related operators see \cite{Nguyen:2021naa}.}. 
\be
Z[b_{\sM_2}] = \langle G_{g}(\sM_{2}) \cdots \rangle
\ee
To see that this is topological, let us consider trying to deform the surface. Most of the action is in two directions transverse to the defect, so let us focus on those, as in Figure \ref{fig:plaquetteBfield}. Note that if we do the following change of variables in the picture:
\be
U_{4} \to g^{-1} U_{4} \label{latmanip} 
\ee
it has the effect of setting $b_{\Box_1} = \mathbf{1}$ but setting $b_{\Box_2} = g$; thus it has moved the surface operator over by one site, but leaves the partition function invariant (as it is just a change of variables in the path integral). By performing operations of this form one can deform the operator however we like; thus there exists a topological operator living on 2d surfaces in the dual lattice. Zooming out to long distances the distinction between dual and original lattices goes away and this becomes a smoothly deformable continuum operator. 

\begin{tcolorbox}
{\bf Exercise:} Prove that the linking of the lattice Gukov-Witten surface operator $G_{g}(\sM_2)$ with the fundamental Wilson line \eqref{latticeWilson} satisfies the desired linking equation \eqref{wilsonlink}. 
\end{tcolorbox}

The existence of a topological operator $G_{g}(\sM_2)$ is a statement that can easily be lifted to the continuum. 
In fact this is a consequence of a larger invariance of the partition function in the presence of the source $b$. Consider a $\mathbb{Z}_N$ valued function $g_{l}$ on the links $l$. Assume that this acts on the dynamical link variables $U_{l}$ and the plaquette variables $b_{\Box}$ as follows:
\be
U_{l} \to U_{l}' = U_{l} g_{l} \qquad b_{\Box} \to b_{\Box}' = b_{\Box} \prod_{l \in \Box} g_{l}^{\dagger}
\ee
(where the latter product is again taken in an oriented fashion). This leaves the twisted plaquette product $W_{\Box}$ -- and thus the whole action $S[U;b]$ -- invariant. The partition function is thus invariant under
\be
Z[b'] = Z[b]
\ee

This invariance is equivalent to the existence of the $\mathbb{Z}_N$ 1-form global symmetry. Note that it is a discrete version of the sort of invariance we saw for the $U(1)$ Maxwell theory around \eqref{maxsource}. Importantly, it {\it not} the same as the $SU(N)$ gauge redundancy discussed around \eqref{latticegauge}, which involved an $SU(N)$ group element on each lattice {\it site}, not a $\mathbb{Z}_N$ on the links. We will shortly see that one can preserve the gauge redundancy while breaking the 1-form global symmetry. 

\subsubsection{Confinement and the 1-form symmetry}
Having now carefully understood that pure $SU(N)$ gauge theory has a $\mathbb{Z}_N$ symmetry under which the fundamental Wilson loop is charged, we can ask about the different phases:
\ben
\item {\bf Confined phase}: here the 1-form symmetry is {\it unbroken}, and the Wilson loop behaves as an area law: $\vev{W_{F}(C)} \sim \exp(-T \mbox{Area}(C))$. The physics interpretation is that the Wilson loop creates a non-Abelian flux tube which has a finite tension $T$; it is these flux tubes which the 1-form symmetry counts (mod N). This is the phase where we normally end up for non-Abelian gauge theory. 
\item {\bf Deconfined phase}: here the 1-form symmetry is {\it spontaneously broken}, and the Wilson loop behaves as a perimeter law $\vev{W_{F}(C)} \sim \exp(-T L[C])$. The flux tubes have melted and condensed. 
\een
Note that from this point of view the confinement/deconfinement transition is philosophically the same as the phase transitions that we know from the ordinary Landau paradigm, just involving a 1-form symmetry rather than a 0-form one! 

We can now ask the following question: what happens if we add matter to the theory, e.g. if we add a scalar charged in the fundamental under $SU(N)$:
\be
S[A,\phi] = \int d^4x \le(\frac{1}{2g^2}\Tr(F_{\mu\nu} F^{\mu\nu}) + (D\phi)^{\dagger} D\phi + m^2 \phi^{\dagger}\phi\ri)
\ee
where $ D_{\mu}\phi = \p_{\mu} \phi + i A_{\mu}^a t^a \phi$ with $t^{a}$ the fundamental representation matrices. 

In this case the 1-form symmetry is explicitly broken, i.e. the putative Gukov-Witten surface operator $G_{g}(\sM_2)$ is no longer topological. 

We can see this in the continuum by noting that the quanta of the $\phi$ field trace out effective worldlines in spacetime, where the coupling of each worldline to the gauge field is given by a fundamental Wilson loop $\vev {W_{f}(C)}$; thus $G_{g}(\sM_2)$ now feels a phase from the {\it dynamical} degrees of freedom, spoiling its topological-ness.  

More concretely, we can also see this on the lattice. To add fundamental matter to the Wilson lattice action, we consider a field $\phi_i \in \mathbb{C}^N$ living on each site $i$ of the lattice. Under the gauge transformation \eqref{latticegauge} we take $\phi_i$ and $U_{\vev{ij}}$ to transform as 
\be
U_{\vev{ij}} \to \Lambda_{i} U_{\vev{ij}} \Lambda_{j}^{\dagger} \qquad \phi_i \to \Lambda_i\phi_i
\ee
so that the coupling $\phi_i^{\dagger} U_{\vev{ij}} \phi_j$ is a gauge-invariant lattice kinetic term for $\phi$. We can now consider the Wilson action with fundamental matter:
\be
S[U,\phi] = -\frac{1}{g^2} \sum_{\Box}\le(W_{\Box} + W_{\Box}^{\dagger}\ri) + t \sum_{\vev{ij}} \phi_i^{\dagger} U_{\vev{ij}} \phi_j \label{gaugewithmat} 
\ee 
(We can also add a further on-site mass term which we neglect for simplicity). If we now try to construct the topological Gukov-Witten operator through the arguments above, we will see that it is simply not possible; the existence of the $\phi$'s renders impossible the manipulations in \eqref{latmanip}. 

This means that once fundamental charged matter exists, there is no longer a clean symmetry-based distinction between the confined and deconfined phases; in all phases we will find that the Wilson law exhibits a perimeter law. (Physically, if you try to create an extended flux tube, it is now allowed to {\it break} into two pieces with a $\phi$ particle ending each tube). Thus there is no order parameter for confinement in this case, because there is no global symmetry. (Note that QCD also has matter in the fundamental representation, and thus does not permit a simple order parameter for confinement.). 

\subsection{Gauge theory: $\mathbb{Z}_2$}
A careful reader will have noticed above that most of the structure of the non-Abelian gauge symmetry played essentially no role in this discussion. One might then reasonably ask: what is the {\it simplest} theory that has the structure above, i.e. that of a discrete 1-form symmetry which could be spontaneously broken? 

The simplest is the $\mathbb{Z}_2$ lattice gauge theory. This is essentially the same as above, except that instead of having a $SU(N)$ element on each link we have a $\mathbb{Z}_2$ element, i.e. a single sign which can be either positive or negative:
\be
a_l = \pm 1
\ee
The analogue of the field strength is also valued in $\pm 1$:
\be
W_{\Box} = \prod_{l \in \Box} a_l
\ee
The action is simply
\be
S[a] = -\beta \sum_{\Box} W_{\Box} \qquad Z = \sum_{\{a_l\}} \exp(-S[a])
\ee
There is a gauge redundancy given by 
\be
a_{\vev{ij}} \to a_{\vev{ij}}' = \Lambda_{i}a_{\vev{ij}}\Lambda_{j} \label{Z2gauge} 
\ee
where $\Lambda_{i}$ is a $\mathbb{Z}_2$-valued field defined on site. 

Importantly, this theory also has a $\mathbb{Z}_2$ 1-form global symmetry. The charged line operator is the {\it Wegner-Wilson line operator}, given by by $W[C] = \prod_{l \in C} a_{l}$. The co-dimension $2$ charge defect -- i.e. the analogue of the Gukov-Witten surface operator -- can be constructed by an argument similar to that above, and turns out to take the explicit form:
\be
G(S) = \prod_{\Box \in S} \exp(-2\beta W_{\Box})
\ee
where again the product over plaquettes is taken over those that intersect the surface in the dual lattice (here the factor of 2 comes from the fact that we are changing the sign of the appropriate plaquette term in the action). 

This theory has a confined phase (unbroken 1-form symmetry) at $\beta \approx 0$ and a deconfined phase (spontaneously broken 1-form symmetry) at $\beta \gg 1$, where the precise location of the critical point can easily be found by numerical simulation. Much more can be said about this, and we will just make a few more quick remarks.

\subsubsection{Kramers-Wannier duality and the Ising spin model} \label{sec:ising} 
In 3d, this theory is dual by Kramers-Wannier duality \cite{Wegner1971} -- i.e. a non-local rewriting of the theory -- to the regular 3d Ising spin model, which has action
\be
S = -\tilde{\beta}\sum_{\vev{ij}} \sig_i \sig_j \label{ising} 
\ee
where $\sig_i = \pm 1$, and where we have $\tilde{\beta} = -\ha \log (\tanh \beta)$. 

Physically speaking, this mapping takes the worldsheets of 2d flux tubes in the gauge model to domain walls that separate up from down spins in the spin model.

The global symmetries may be confusing here, as clearly the model \eqref{ising} has a $0$-form $\mathbb{Z}_2$ global symmetry $\sig_i \to -\sig_i$, and not a 1-form global symmetry. The precise statement of Kramers-Wannier duality for the 3d Ising model is that the $\mathbb{Z}_2$-invariant sectors are mapped into each other, and a genuine charged operator on one side is mapped into a non-genuine charged operator (i.e. something with a tail attached) on the other. 

Of course, this is a very well-studied model for magnetism. The action penalizes configurations where $\sig_i$ differs from its neighbours, and thus the low-temperature phase $(\tilde{\beta} \to \infty)$ has all of the $\sig_i$ with the same sign, and the $\mathbb{Z}_2$ spin symmetry spontaneously broken, the {\bf ferromagnetic} phase. As we increase temperature (i.e. take $\tilde{\beta} \to 0$) we eventually hit a phase transition at $\tilde{\beta} \approx 0.221$\footnote{See \cite{Xu:2018hwn} for a recent measurement of this critical coupling from Monte Carlo simulations.} and the model enters a disordered {\bf paramagnetic} phase. The continuous critical point separating these two -- the 3d Ising transition -- is a strongly coupled fixed point about which we nevertheless know a lot. Currently the most precise estimates of the critical exponents come from the conformal bootstrap \cite{El-Showk:2012cjh} (see \cite{Poland:2018epd} for a review). 

\subsubsection{Higgs/confinement continuity} Let us study slightly further this model in 3d. We can imagine coupling in $\mathbb{Z}_2$ matter to this gauge field, i.e. introduce $\phi_i = \pm 1$, a $\mathbb{Z}_2$-valued field on the {\it sites} who transforms under the gauge transformations as
\be
\phi_i \to \Lambda_i \phi_i 
\ee
We then have the following model, the $\mathbb{Z}_2$ analogue of \eqref{gaugewithmat}:
\be
S[a,\phi] = -\beta \sum_{\Box} W_{\Box} - J \sum_{\vev{ij}} \phi_i a_{\vev{ij}} \phi_j \label{Z2matter} 
\ee
This model illustrates some important physical points:
\ben
\item First, by arguments identical to those around \eqref{gaugewithmat}, the presence of matter breaks the $\mathbb{Z}_2$ 1-form symmetry; thus at generic points in the $(\beta, J)$ space the model has no global symmetries. 

\item Here $J$ controls the matter fluctuations and $\beta$ the gauge fluctuations. What is the phase diagram as a function of $J$ and $\beta$? If we set $J \to 0$ then there is a deconfined phase at larger $\beta$ and a confined phase at small $\beta$. 

Let's now consider the limit where $\beta \to \infty$; this essentially suppresses all the gauge field fluctuations, so we expect to find a setup where $W_{\Box} = 1$ on all plaquettes, which means that $a_{\vev{ij}} = 1$ (or its gauge-equivalent) on all links. We thus end up with the ordinary Ising spin model \eqref{ising} expressed in terms of the $\phi_i$; as described above this model has a phase transition at some critical point $J^*$. 

Note that this phase transition is -- morally -- condensing the $\phi_i$ fields. So in conventional language, we would call it the ``Higgs'' phase. We say morally because the $\phi_i$ are gauge charged and thus we cannot really speak about a condensate of them, but it is a fact that the phase transition happens.  

 \begin{figure}[h]
\begin{center}
\includegraphics[scale=0.6]{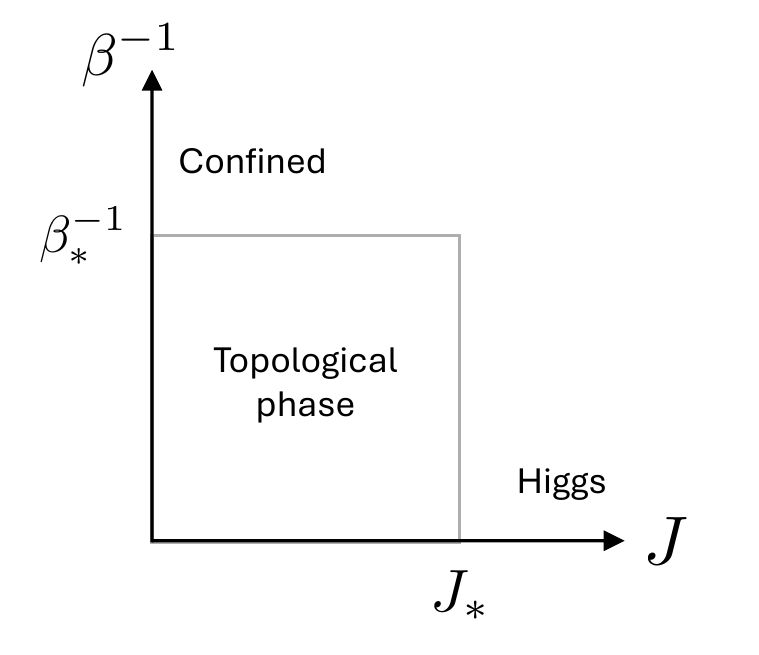}
\end{center}
\caption{Cartoon phase diagram of model described by \eqref{Z2matter}; note that one can move continuously from the Higgs to the confined phase. Details of the physics along the diagonal have been omitted, see \cite{Somoza:2020jkq} for a discussion.} 
\label{fig:Z2diag}
\end{figure}

\item The phase diagram of this model, as shown in Figure \ref{fig:Z2diag} is now somewhat interesting. Note that there are basically two phases; one that is continuously connected to the deconfined phase of the gauge theory -- we will call this the {\bf topological} phase -- and another, which is continuously connected both to the Higgs phase on one side, and the confined phase on the other. If nothing else happens, this implies the baffling statement that the Higgs and confined phases are {\bf continuously connected} \cite{PhysRevD.19.3682}. 

From the point of view of gauge theory as formulated as a perturbation theory in the continuum, this statement is very surprising. However it is indeed true; the point is simply that there are no global symmetries that can be used to diagnose the phases, and thus, there is nothing {\it stopping} the Higgs and confined phases from being connected to each other. The status of Higgs/confinement continuity in more complicated theories -- such as, e.g. QCD -- is a matter of active study (see e.g. \cite{Dumitrescu:2023hbe,Cherman:2024exo}). 

\item It is not an accident that the phase diagram appears symmetric with respect to reflection about the diagonal. There is in fact a self-duality symmetry that switches the two sides; see \cite{Somoza:2020jkq} for a discussion of the rich physics appearing at the phase transition at the self-dual point. As of this writing, there is still not yet a useful continuum description for this phase transition, which is possibly the simplest beyond-Landau critical point.

\een

\subsection{Effective description of spontaneously broken $\mathbb{Z}_N$ 1-form symmetry: topological order} \label{effdesc} 
We now turn to a different question: suppose we have a theory that has a spontaneously broken $\mathbb{Z}_N$ 1-form symmetry, such as deconfined $SU(N)$ gauge theory, or maybe the deconfined phase of the $\mathbb{Z}_2$ gauge theory described above. What is the {\it effective description} of the long distance physics? (Here we will be motivated by $SU(N)$ gauge theory and its associated $\mathbb{Z}_N$ 1-form symmetry). 

We should first ask: what are the long-distance degrees of freedom? Generically, there will be a gap to all local degrees of freedom, but we should be able to access objects corresponding to the fundamental Wilson line and the Gukov-Witten surface operator, which have a non-trivial braiding phase when they link as in \eqref{wilsonlink} 
\be
G(S^2) W(C) = e^{\frac{2\pi i}{N}}W(C) \label{wilsonlinkrpt} 
\ee
where here we have omitted the subscripts and are now referring to the minimal charge object of each class. Note that though $G(S^2)$ was topological by construction, in the extreme infrared limit $W(C)$ is {\it also} topological -- this is because it has vanishing correlation with all other operators (which themselves have only exponentially suppressed correlations with each other), but it is nevertheless not zero (as it would have been if it had an area law). 

\subsubsection{BF theory}
We thus seek a topological field theory to capture the physics of this braiding. Our discussion will be in 4d with an eye towards the $SU(N)$ gauge theory, but by modifying a few indices here and there it works for $\mathbb{Z}_N$ p-form symmetry in any number of dimensions. 

The required object is called {\bf BF theory}, and can be defined in terms of a local action involving a 1-form and 2-form gauge field $A_1$ and $B_2$ respectively: 
\be
S[A, B] = \frac{i N}{2\pi} \int_{\sM_4} B \wedge dA = \frac{i N}{2\pi} \int_{\sM_4} A \wedge dB \label{bf} 
\ee
This is a topological field theory with a long history \cite{horowitz1989exactly}; see \cite{Banks:2010zn,Maldacena:2001ss} for discussions in the high-energy literature. It is a kind of mixed, higher-form Chern-Simons theory. The metric does not appear in its construction: it does not care about distances, and cares only about topology. As we will see, it can also be thought of as a gauge theory with a $\mathbb{Z}_N$ gauge group. 

This action is invariant under the following gauge redundancies:
\be
B_2 \to B_2 + d\Lambda_1 \qquad A_1 \to A_1 + d \lam_0 \ . \label{bfgauge}  
\ee
It's important that both of these are compact gauge fields, i.e. the gauge transformations are valued in $U(1)$ and just as around \eqref{Dirac} the associated monopole numbers are quantized appropriately, i.e. if we have $F = dA$ and $H = dB$ then over all 2 and 3-cycles respectively we have:
\be
\oint_{\sM_2} F_2 = 2\pi\mathbb{Z} \qquad \oint_{\sM_3} H_3 = 2\pi\mathbb{Z} 
\ee
This imposes a restriction on the parameter $N$. Consider studying the action the theory on $S^1 \times S^{3}$. Now consider doing a  {\bf large gauge transformation} by gauge parameter $e^{i\lambda}$, where $\lambda$ winds once around the $S^1$, i.e. if the $S^1$ coordinate is $\phi \in [0,2\pi]$
\be
\lambda = \phi \ . 
\ee
Here the word ``large'' refers to the fact that this gauge transformation cannot be continuously deformed to the identity.\footnote{Note that in the literature a ``large gauge transformation'' can also sometimes refer to a gauge transformation on a non-compact space, where the gauge parameter does not die away at infinity. This is a different situation, as those large gauge transformations typically {\it can} be continuously connected to the identity. They are also (usually) actually not ``gauge'', in they generate global symmetries, i.e. actual physical invariances of the system, unlike the large gauge transformations discussed above, which {\it are} redundancies of the description.} 

The gauge transformation $e^{i\lambda}$ is single-valued on the circle, and thus this is a perfectly valid gauge transformation to do. Under this transformation $A \to A' = A + d\lambda = A + d\phi$
\be
S[A',B] - S[A,B] = \frac{i N}{2\pi} \int_{S^1 \times S^3} d\phi \wedge H = \frac{i N}{2\pi} \le(\int_{S^1} d\phi\ri)\le(\int_{S^3} H\ri) = 2\pi i N q
\ee
where $q$ is the integer-quantized ``monopole number'' of the integral of $H$ over $S^3$. Note the horrifying fact that the action is {\it not} invariant under this large gauge transformation. We are saved by the fact that we do not study the action itself; rather this is an intrinsically quantum theory which makes sense only when we insert the action into a path integral as
\be
Z = \int [dA] [dB] \exp(-S[A,B])
\ee
and thus what we actually need is for the {\it exponential} of $S$ to be invariant. This will be the case provided we have
\be
N \in \mathbb{Z} \ . 
\ee
meaning that the shift in $S$ affects the partition function by an invisible factor of $e^{2\pi i}$. 

We thus see that the theory defined by \eqref{bf} has its parameter $N$ quantized. (An alternative classic argument involving defining Chern-Simons theory as the boundary of a higher-dimensional manifold can be found in \cite{Witten:2003ya}). This rigidity of the theory is generally the case for topological field theories, which are discrete objects that do not permit continuous deformations. 

We now study the dynamics. The local equations of motion are
\be
dA = 0 \qquad dB = 0
\ee
and thus there are no local degrees of freedom. However there are extended operators we can insert on closed 1-cycles $C_1$ and 2-cycles $\sM_2$ respectively:
\be
W(C_1) = \exp\le(i\oint_{C_1} A\ri) \qquad G(\sM_2) = \exp\le(i\oint_{\sM_2} B\ri)
\ee
These objects are invariant under the gauge transformations \eqref{bfgauge}. We now make the main dynamical statement -- in BF theory these two objects have a nontrivial $\mathbb{Z}_N$ braiding, i.e.
\be
\big\langle W(C_1) G(\sM_2) \big\rangle \sim \exp\le(\frac{2\pi i}{N} L(C_1, \sM_2)\ri) \label{BFlinking} 
\ee
where $L$ is the {\it linking number} of $C_1$ and $\sM_2$, a well-defined object in four dimensions. 

The proof of this is not completely obvious, and is a good exercise. 

\begin{tcolorbox}
{\bf Exercise:} Prove the linking relation \eqref{BFlinking}. {\bf Hint:} consider the equations of motion with the surface operators inserted and study the on-shell action for a simple configuration of $C_1$ and $\sM_2$. 
\end{tcolorbox} 
A solution to the exercise is provided below. Insert the two surface operators into action to find:
\be
S[A,B] = \frac{i N}{2\pi} \int B \wedge dA +  i\oint_{C_1} A + i \oint_{\sM_2} B
\ee
Varying with respect to $A$ and $B$ we find that the equations of motion become
\begin{align} 
\frac{N}{2\pi} dA(x) & = -\delta_{\sM_2}(x)  \label{BFeom} \\
\frac{N}{2\pi} dB(x) & = -\delta_{C_1}(x) \ . 
\end{align} 
Let us denote the solutions to this set of equations to be $\overline{A}$, $\overline{B}$. Plugging this in to find the on-shell action we find that two terms cancel and we are left with: 
\be
S[\overline{A}, \overline{B}] = i \oint_{C_1} \overline{A}
\ee
In other words, we should determine the $\overline{A}$ sourced by the insertion of $G(\sM_2)$ through \eqref{BFeom} and evaluate it on the location of the insertion of $W(C_1)$. 

Let us now assume that $C_1$ links $\sM_2$ once. The geometry is the same as in Figure \ref{fig:goldstone1form}. Consider integrating \eqref{BFeom} over a small ball $B_2$ which is bounded by $C_1$; this intersects $\sM_2$ at a point and we thus find that
\be
\frac{N}{2\pi} \int_{B^2} d\overline{A} = \frac{N}{2\pi} \oint_{C_1} \overline{A} = -1
\ee
Plugging everything in, we find that in the case where the two manifolds link once we have:
\be
S[\overline{A}, \overline{B}] = -\frac{2\pi i}{N}
\ee
From this it is clear that if they had linked multiple times we would obtain an extra factor on the right-hand side. On the other hand, if they don't link at all we obtain zero; we have thus found the relation \eqref{BFlinking} (up to an overall sign which depends on a choice of orientation). 

Thus we see that we have realized the desired linking algebra \eqref{wilsonlinkrpt}. A phase described by such a topological field theory is usually called an example of {\bf topological order}, and is often exhibited as an example of a {\it beyond-Landau} phase of matter \cite{wen1990topological,wen2004quantum}. This name predates the technology of higher-form symmetry: in this section we have just shown that the spontaneous breaking of a 1-form symmetry is an example of topological order, and thus can be understood from the point of view of an enlarged Landau paradigm involving higher-form symmetry. 

\subsubsection{Ground states in a topologically ordered phase}
If we spontaneously break a conventional 0-form $\mathbb{Z}_N$ symmetry, we have $N$-fold degenerate ground states which are acted upon by the action of the symmetry. They are precisely degenerate in the limit of infinite volume. 

Here we will discuss the higher-form analogue of this phenomena; again if we break a $1$-form $\mathbb{Z}_N$ symmetry, we will have degenerate ground states. In this case their number is determined by the {\bf topology} of the spatial manifold. 

As an example, let us consider quantizing the theory on $\mathbb{R} \times S^2 \times S^1$, where the first $\mathbb{R}$ is ``time''. (Note that this is the first time in these notes that we have worked in a Hamiltonian framework). It is instructive to quantize the theory carefully. We just note the following points, whose verification is left as an exercise: 

\ben
\item There are no local degrees of freedom. There may still be ``delocalized'' states that live on the whole spatial manifold. 
\item The Hamiltonian is zero, so all of these states must have vanishing energy: they are  all ``ground states''. 
\item The only non-trivial operators that act on the space of states are the following
\be
W \equiv \exp\le(i \oint_{S^2} B\ri) \qquad G \equiv \exp\le(i\oint_{S^1} A\ri) 
\ee
These are the only topologically non-trivial operators that we can construct. As the equations of motion set $dA = dB = 0$ any topologically trivial operators can be shrunk to zero and collapsed, and have no action on any state. 

Note also that these are the objects that we saw had a non-trivial braiding in Euclidean space \eqref{BFlinking}. Here we are interpreting them as quantum operators acting on the Hilbert space, i.e. they act at a given time. One can directly show from the canonical commutation relation of the local fields $A$ and $B$ that the non-local operators satisfy the following equal-time commutation relation:
\be
G W = e^{\frac{2\pi i}{N}} W G, \label{gsalg} 
\ee
\een
We now seek to find the space of states, i.e. we seek to find a representation of the algebra \eqref{gsalg}. There is no one-dimensional representation, and thus the ground state is necessarily not unique. The smallest representation is $N$-dimensional, with basis elements which we label by $\ket{n}, n \in \{0, 1,\cdots (N-1)\}$. The operators act as
\be
W \ket{n} = \ket{(n + 1)\;\mbox{mod } N} \qquad G\ket{n} = \exp\le(\frac{2\pi i n}{N}\ri) \ket{n}
\ee
These $N$ states are the ground states of BF theory on $S^2 \times S^1$. Physically, the state $\ket{n}$ corresponds to having $n$ flux tubes threading the $S^1$; acting with the Wilson line on the $S^1$ creates another flux tube, and $G$ measures the number of tubes (mod $N$). 

Note that this state exists only because of the non-trivial topology; had we played the same game on $S^3$ we would have found a single unique ground state. This is a hallmark of {\bf topological order.} 

\begin{tcolorbox}
{\bf Exercise:} Prove the commutation relation \eqref{gsalg}.
\end{tcolorbox} 

\begin{tcolorbox}
{\bf Exercise:} Consider a situation where there are dynamical string worldsheets -- with tension $T$ -- that couple to the 2-form $B_2$. Compute in some suitable approximation the splitting of the degenerate energy levels we have found above. {\bf Hint:} -- you should find $\Delta E \sim \exp(-4\pi T R^2)$, where the $S^2$ has radius $R$. 
\end{tcolorbox} 

There is much more to say about topological order. We will make only a few more remarks:
\ben
\item A simple and extremely instructive Hilbert space realization of the topological phase of the $\mathbb{Z}_2$ gauge theory in $(2+1)d$ is given by the so-called {\it toric code} \cite{Kitaev:1997wr}.   
\item These lectures were first given at the Jena school for theoretical physics, where the other lecture series was about entanglement entropy. In fact, a system that has topological order will exhibit a signature of it in its entanglement entropy. There is a {\it reduction} of entropy arising from the fact that every one of the un-endable-strings that we discuss that enters a region of space must also exit the same region of space. In the two-dimensional toric code case discussed above this so called ``topological entanglement entropy'' \cite{Kitaev:2005dm,PhysRevLett.96.110405} is a constant contribution which takes the form
\be
S(\mbox{disc}) = \mbox{non-universal area law} - \log 2
\ee
Clearly for a $\mathbb{Z}_N$ gauge theory the $2$ is replaced by an $N$. The higher dimensional case is discussed from the microscopic theory in \cite{Grover:2011fa} and from the effective BF description in \cite{Fliss:2023uiv}\footnote{As of this writing there appears to be a slight discrepancy between these two descriptions.}. As described extensively in \cite{Kitaev:2005dm,PhysRevLett.96.110405}, the $\log N$ contribution may be cleanly separated from the area law by adding and subtracting the entanglement entropy defined on different spatial regions. 
 
 \item The usual first example of topological order in most textbooks is actually usually not the $SU(N)$ gauge theory, but rather the {\it fractional quantum Hall effect}. It turns out that if you confine many electrons to a plane and apply a strong magnetic field, the frustrated dynamics arising from the interaction of Landau levels and Coulomb repulsion results in a topologically ordered state, of which the simplest example be described by the following Chern-Simons action:
 \be
 S[A] = \frac{ik}{8\pi} \int_{\sM_3} A \wedge dA
 \ee
 where $A$ is not the photon of real life, but rather a new emergent gauge field. There is again an emergent $\mathbb{Z}_k$ 1-form symmetry that is spontaneously broken, but its realization is slightly different due to the different structure of the Chern-Simons term. See \cite{Tong:2016kpv} for an entrance to this fascinating field. 
\een

%\subsection{Fractional quantum hall effect} 

\section{Non-invertible symmetries: the axial anomaly}
We will now discuss a different generalization of the idea of symmetry -- {\it non-invertible symmetries}.

 Everything we have discussed so far has involved topological charge defects, whose fusion obeys a group composition law \eqref{groupcomp}:
\be
U_{g}(\sM_{d-1}) U_{g'}(\sM_{d-1}) = U_{g g'}(\sM_{d-1}),
\ee
where $g, g' \in G$ with $G$ some group. In particular, this means that given a particular charge defect labeled by $g$, there is always an inverse element $g^{-1}$ such that
\be
U_{g}(\sM_{d-1}) U_{g^-1}(\sM_{d-1}) = \mathbf{1}
\ee
However one could ask -- is it possible for there to exist topological defects that obey a more general fusion rule, e.g. of the form?
\be
U_{a}(\sM_{d-1}) U_{b}(\sM_{d-1}) = \sum_{c} N^{c}_{ab} U_{c}(\sM_{d-1})
\ee
where, e.g. the right hand side involves a non-trivial sum, rather than just a single element? In that case it is possible that some elements will not have an inverse. This is the idea of {\bf non-invertible} symmetries. 

Examples of this sort of phenomena have been known for a very long time in 2d CFT \cite{verlinde1988fusion}. Their study has exploded recently, with many new and physically relevant examples found in higher dimensions. We will not make an attempt to survey these developments here, but instead refer the reader to recent reviews \cite{Cordova:2022ruw,Schafer-Nameki:2023jdn,Shao:2023gho} which provide an entry point to the literature. 

Before moving on to our main focus, we do however mention one important example, which exists in the familiar 2d Ising model. This theory has an ordinary 0-form invertible $\mathbb{Z}_2$ symmetry that acts on the spins as $\sig \to -\sig$, and this symmetry is associated with an topological line in the manner that is now familiar. However, when tuned to its critical point, the model has an extra topological line which is non-invertible in the sense described above. One way to understand this line is to consider taking half of the 2d space and performing Kramers-Wannier duality on it. As this is a self-duality for the 2d Ising model at the critical point, it will map the bulk of the space to itself. However the junction between the original model and the dual model has some non-trivial object living on it. This turns out to be a topological line which is non-invertible in the sense described above \cite{Frohlich:2004ef,Chang:2018iay}. It is beautifully reviewed from several different points of view in \cite{Shao:2023gho}. 

In these notes we will now study how these symmetries present us with a new way to think about the venerable Adler-Bell-Jackiw anomaly. 
\subsection{Different types of anomalies}
Let us remind ourselves about the physics of anomalies, starting from the example of a massless Dirac fermion in 4d. Consider first the theory of a Dirac fermion alone, which we couple to an {\it external} $U(1)$ gauge field source $a$:
\be
S[\psi;a] = \int d^4x \bpsi \le(\slashed{\p} - i \slashed{a}\ri) \psi
\ee
Let us remind ourselves of the symmetries. The action above is invariant under two possible phase rotations of the Dirac fermion, $U(1)_V$ and $U(1)_A$:
\be
\psi \to e^{\frac{i\al}{2}} \psi \qquad \psi \to e^{\frac{i \al}{2}\ga^5} \psi
\ee
which have corresponding currents
\be
j^{\mu}_V = \bpsi \ga^{\mu} \psi \qquad j^{\mu}_A = \bpsi \ga^5 \ga^{\mu}\psi
\ee
Above we have coupled an external gauge field source to the vector current. Now it is well-known that if we insert this classical action into a path integral as
\be
Z[a] = \int [d\psi d\bpsi] \exp(-S[\psi;a])
\ee
then in the quantum theory with the external source turned on the axial current is no longer conserved \cite{Adler:1969gk,Bell:1969ts}: 
\be
d \star j_{A} = \frac{1}{4\pi^2} f \wedge f \qquad f = da \label{thooft} 
\ee
The result above is one-loop exact. 

Note however that it is not conserved in a {\it very} specific way: the nonconservation of the axial current is {\it precisely} determined by the external source. This is called an {\it 't Hooft anomaly}. 

From the point of view of obtaining predictive power over the theory, a 't Hooft anomaly is generally {\it better} than not having any anomaly at all, as one can obtain non-trivial constraints over the dynamics; after all, we can always choose to set the applied source to zero (so that we have a conserved current again), but we can do more. For example, it is clear that a theory with \eqref{thooft} cannot have a trivial unique gapped ground state, as then one could imagine a setup where the right-hand side of the equation (which we can control freely) is not zero, but the left hand side (which is constrained by the gapped dynamics) is. 

However, in these lectures we will not be discussing 't Hooft anomalies. Instead we want to consider {\it not} the theory above -- where $a$ was frozen -- but rather massless QED. To that end we will promote $a$ to a dynamical gauge field $A$, with its own kinetic term: 
\be
S[\psi,A] = \int d^4x \le(\frac{1}{4 e^2} F^2 + \bpsi \le(\slashed{\p} - i \slashed{A}\ri) \psi\ri) \qquad F = dA,
\ee
and now {\it also} integrate over it in the path integral, i.e.
\be
Z = \int [dA] [d\psi] \exp(-S[\psi,A])
\ee
Let us now revisit the anomaly equation \eqref{thooft}, which now has a very different character:
\be
d \star j_{A} = \frac{1}{4\pi^2} F \wedge F
\ee
as the right-hand side is now a dynamical operator, which fluctuates. Thus it now naively seems like we have lost the symmetry entirely -- the current is just not conserved any more. We will refer to this kind of anomaly -- where the right-hand side is an operator -- as an {\it Adler-Bell-Jackiw} anomaly\footnote{This  nomenclature seems a little imperfect, but I believe it is now standard.}. 

Let us nevertheless try to make sense of it. The first obvious thing to do is consider defining the following shifted current:
\be
\star \tilde{j}_A = \star j_{A} - \frac{1}{4\pi^2} A \wedge F
\ee
This satisfies $d\star \tilde{j}_A = 0$. However it does not actually make sense as a local operator: it is not gauge invariant under small gauge transformations $\delta A = d\Lambda$. 

However we could try to ignore this, and instead simply integrate the whole thing over a 3-manifold anyway to try and make a conserved charge defect, i.e. imagine constructing the following object:
\be
\hat{U}_{\al}(\sM_3) = \exp\le(i\frac{\al}{2} \int_{\sM_3} \le(\star j_{A} - \frac{1}{4\pi^2} A \wedge dA\ri)\ri) \label{illdef} 
\ee
Because we are considering an integral, this is now gauge-invariant under small gauge transformations $\delta A = d\Lambda$. It is also invariant under small deformations of $\sM_3$ and naively it looks like it forms a topological charge defect; the first term rotates an operator with integer axial charge $q$ by $\sO \to e^{\frac{i q \al}{2}}\sO$.

The issue however is invariance under {\it large} gauge transformations. Recall from Section \ref{effdesc} that this quantizes the coefficients of Chern-Simons terms in the action; if we want consistency of this object in a situation where $\sM_3$ has a non-trivial $1$-cycle, then we require that 
\be
\al \in 2\pi\mathbb{Z}
\ee
This means that we can no longer perform a continuous rotation, i.e. it looks like there is trivial action on all operators \footnote{There is a possible $\mathbb{Z}_2$ action depending on whether $q$ is even or odd; this $\mathbb{Z}_2$ is essentially $\psi \to -\psi$, which is part of the $U(1)$ gauge symmetry and is not anomalous.}. 

Thus we cannot form a topological operator on general 3-manifolds $\sM_3$, and it really does look like we have lost the axial symmetry. This may feel somewhat peculiar. In particular, note that the left hand side has a universal description in terms of the current for the (preserved) 1-form symmetry $U(1)^{(1)}_m$, and one might hope that this could be given a precise characterization in terms of some sort of intertwining of these symmetries. 

Note also that if we were willing to restrict our attention to topologically trivial 3-manifolds such as $\mathbb{R}^3$ then technically speaking one could get away with ignoring the issue of large gauge transformations; nevertheless we might worry that we are missing some key physics. 
\subsection{Charge defect operators for $U(1)_A$}
The situation changed when the works \cite{Choi:2022jqy,Cordova:2022ieu} explained how to construct topological charge defects for the ABJ anomaly. Let us begin by considering only rotation angles that are simple fractions of the form 
\be
\al = \frac{2\pi}{N}
\ee
where $N \in \mathbb{Z}$, which is an arbitrary integer. For the reasons above $\hat{U}_{\al}$ does not exist. However let us now consider a different operator:
\be
D_{\frac{1}{N}}(\sM_3) \equiv \int [dB] \exp\le[i\int_{\sM_3} \le(\frac{2\pi}{2N} \star j_A + \frac{N}{4\pi} B \wedge dB + \frac{1}{2\pi} B \wedge dA\ri)\ri] \label{noninvdef} 
\ee
Here we have done something radical -- we have introduced a new dynamical compact gauge field $B$ and given it a topological Chern-Simons action with coefficient $N$, and then we have coupled it to the dynamical photon $A$. We have then done a path integral over that field. 

Let me now state some properties of this object. From the first term, it generates a phase rotation on local operators with axial charge as $\sO \to e^{\frac{2\pi i q}{2N}} \sO$. It is perfectly well-defined and gauge-invariant since $N$ is in the numerator everywhere. However if we now try to impose the equations of motion of $B$ we find the following curious equation:
\be
B = -\frac{A}{N} \label{fraca} 
\ee
If you eliminate $B$ using this relation and plug it into the action \eqref{noninvdef} then you recover -- from a well-defined starting point -- the previously {\it ill}-defined object \eqref{illdef} with a fractional rotation angle $\hat{U}_{\frac{2\pi}{N}}$! 

Thus we have succeeded in obtaining a well-defined topological charge defect that generates a non-trivial rotation, at the cost of introducing a new field $a$ on the defect (or equivalently, coupling a TQFT to the bulk degrees of freedom). 

Note that \eqref{fraca} does not quite make sense -- if you integrate $d$ of it over a non-trivial cycle you find improperly quantized fluxes. Thus there is a more rigorous way then what we have just done to show that the defect is topological, involving gauging a $\mathbb{Z}_N$ subgroup of $U(1)_m^{(1)}$ on half of the space, but the intuitive argument above gives the correct answer. 

Let us now explore the consequences of this new degree of freedom. Let's consider fusing two of these objects with opposite rotation angles:
\be
D_{\frac{1}{N}}(\sM_3) D_{-\frac{1}{N}}(\sM_3) = \int [dB_1] [dB_2] \exp\le[i\int_{\sM_3}\le( \frac{N}{4\pi} B_1 \wedge dB_1 - \frac{N}{4\pi} B_2 \wedge dB_2 + \frac{1}{2\pi} (B_1-B_2) \wedge dA\ri)\ri]
\ee
This object is {\bf not} the identity operator, as it would have been for a normal symmetry -- note that the difference gauge field $B_{\Delta} \equiv B_1 - B_2$ has a nontrivial coupling to the photon magnetic field. (In fact the right hand side is a kind of topological defect called a {\it condensation defect}.   \cite{Roumpedakis:2022aik,Choi:2022zal}). 

Thus we cannot invert this symmetry any more, and we see that the true effect of the ABJ anomaly is to turn the ordinary axial symmetry into a {\bf non-invertible symmetry}; there is a topological defect associated with it which counts the charge, but the action of performing a charge rotation can no longer be undone. 

How do we use this object? Its action on normal local operators is conventional and invertible, and as such it can be used to derive selection rules on scattering amplitudes and constrain the terms that can appear in RG. This works the normal way on $\mathbb{R}^4$ but can result in interesting subtleties for other manifolds; see \cite{Cherman:2022eml} for a recent discussion. 

However its action on non-local objects such as 't Hooft lines is different. In general if one drags the 't Hooft line through one of these objects it acquires a {\it tail}, i.e. it becomes the boundary of a two-dimensional topological surface which is $\exp\le(\frac{i}{N}\int_{S} F\ri)$. This can be understood as a manifestation of the Witten effect \cite{Witten:1979ey}. As a reminder, the Witten effect tells us that if one introduces a topological coupling $\th$ as $\th \int_{\sM_4} F \wedge F$ -- which is essentially what we are doing when we perform the axial rotation above -- then this binds an fractional electric charge charge proportional to $\th$ to a magnetic monopole. A fractional electric Wilson line is not a well-defined object; the closest one can get to it is to integrate $F$ over a two-surface $S$ whose boundary can now be imagined to be the Wilson line. This explanation has been very telegraphic, see \cite{Choi:2022jqy} for more details. 

Finally, you might feel worried that we have only constructed a rotation angle by $\frac{2\pi}{2N}$, for any $N$; however clearly by composing this $p$ times we can obtain any {\it rational} fraction $\frac{p}{N}$ of $2\pi$. (In fact there is a simpler route which uses a different and slightly more involved TQFT from the simple Chern-Simons theory to arise at the same rational covering of the unit circle). 

The existence of rational numbers may seem peculiar. There is an alternative route to constructing the charge defects \cite{GarciaEtxebarria:2022jky,Karasik:2022kkq}; rather than introducing a Chern-Simons field on the defect, you introduce a compact phase field $\th(x)$ and consider the following gauge invariant object, parametrized by a continuous $U(1)$ element $\al$ (and not a rational number).  
\be
U_{\al}(\sM_3) = \int [d\th] \exp\le(i\frac{\al}{2} \int_{\sM_3} \le(\star j_{A} - \frac{1}{4\pi^2} (A - d\theta) \wedge dA\ri)\ri) \label{newdef} 
\ee
Here under a $U(1)$ gauge transformation the scalar $\theta$ transforms as:
\be
A \to A + d\lambda \qquad \theta \to \theta + \lambda
\ee
This object has the advantage\footnote{It also has a disadvantage: ideally one would like to be able to orient these charge defects with one dimension in {\it time} and then quantize the theory on an equal-time slice that intersects the defect. In that case one usually obtains a finite Hilbert space on the defect interacting with the regular quantum field theory degrees of freedom. The object \eqref{newdef} seems to have an infinite-dimensional Hilbert space living on it and is thus somewhat ill-defined in comparison to the mathematically precise degrees of freedom that constitute \eqref{noninvdef}. There is something to be resolved further here; I am grateful to S-H. Shao for discussions on this point.}   that it allows one to construct a local density on the defect
\be
\star \tilde{j}_A =  \star j_{A} - \frac{1}{4\pi^2} (A - d\theta) \wedge dA
\ee
As this object requires the phase field $\theta$ it cannot be defined {\it off} the defect. The existence of this density permits one to prove an analogue of Goldstone's theorem for this non-invertible symmetry, which can be viewed as a perhaps unnecessarily fancy way to understand why $U(1)$ axions are massless (or, in real life, why the $\pi_0$ is so light). Very recently other approaches to constructing a full $U(1)$ symmetry have been constructed \cite{Arbalestrier:2024oqg}.

\section{More applied applications}
Here we turn to other applications of higher-form symmetries. Our considerations up till now have been somewhat formal: though we have studied some theories which exist in real life, we have not studied them under very realistic situations. As we will see below, it is in fact quite interesting to consider the realization of these higher-form symmetries at {\it finite temperature} $T$, providing us with very interesting connections to magnetohydrodynamics and astrophysics. In this section we assume the reader is familiar with the approach to finite-temperature quantum field theory in the formalism of the Euclidean path integral, as described for example in the standard textbooks \cite{kapusta2007finite,le2000thermal}. 
%\subsection{Mean string field theory}
\subsection{Finite temperature: static phenomena}
 A very physically relevant example is to study hot QED coupled to massive electrically charged matter (as in Section \ref{sec:matter}), i.e.
\be
S[\phi, A] = \int d^4x\le(\frac{1}{4e^2} F^2 + (D\phi)^{\dagger} D\phi + m^2 \phi^{\dagger}\phi + \cdots \ri)
\ee
If we are interested in static phenomena, we can understand the physics at finite $T$ by placing the system on $S^1 \times \mathbb{R}^{3}$, where the $S^1$ has length $\beta = \frac{1}{T}$. Note that the covariant derivative acts as $D_{\mu} \phi = \p_{\mu} \phi - i q A_{\mu} \phi$. 

Recall that there is generically only a single magnetic 1-form symmetry with current $J^{\mu\nu} = \ha \ep^{\mu\nu\rho\sig} F_{\rho\sig}$. Upon dimensional reduction, from the 3d point of view this will break into a $0$-form symmetry $U(1)^{(0)}_m$ with current $j^{i} = J^{i\tau}$ and a $1$-form symmetry $U(1)^{(1)}_m$ with current $J^{ij}$. We should now ask -- as usual -- how these symmetries are realized on the $\mathbb{R}^{3}$, i.e. the non-compact directions. 

To understand this, let us try and compute an effective action. Let's decompose the vector potential into $A_{\mu} = (A_{\tau}, A_{i})$ and understand what each of the components {\it feels}, by integrating out the bosons at finite temperature. Let us attempt to understand the expected result from elementary physics, which is that of a gas of charged bosons at finite temperature, interacting with electric fields. In quantum statistical mechanics, we often consider bosons at finite chemical potential $\mu$, which can be understood as the background value of the real-time vector potential $A_{t} = \mu$. Various quantities typically depend on $\mu$ though a Bose-Einstein distribution of the form
\be
n(E) = \frac{1}{e^{\beta (E \pm q\mu)}-1} 
\ee
where the sign depends on whether we are considering the particle or the antiparticle, which have opposite charge. Now when we go to Euclidean time we have $A_{\tau} = i A_t = i \mu$; let's also consider a limit where the mass of the particle is very high, so that $E \approx m$ (i.e. we are ignoring the contribution from the motion of the particles), and also that $\beta \gg E^{-1}, \mu$ (so that the mass and $\mu$ are high compared to the temperature). 

In that case this distribution takes the form
\be
n(E) \sim e^{-\beta m \pm i q \beta A_{\tau}}
\ee
Thus if we now imagine constructing an effective action for $A_{\tau}$, the ingredients take the form shown above, and thus in the limit shown we expect an answer of the form: 
\be
S[A_{\tau}] = \beta \int d^3x \le(\frac{1}{2e^2} (\p_{i} A_{\tau})^2 + c_1 e^{-\beta m} \cos (q \beta A_{\tau}) + \cdots \ri) \label{effacfiniteT1} 
\ee
where the first term comes from the bare kinetic term, we have added together the particle and anti-particle contributions and $c_1$ is a non-universal constant. Note that the answer is periodic in $q \beta A_{\tau}$, as it must be since a background of the form $q\beta A_{\tau} = 2\pi \mathbb{Z}$ around a non-contractible $S^1$ is gauge-equivalent to no background at all. 

The reader who is happy with this intuitive argument is free to move on; however the reader who wants a more concrete derivation may want to derive this result more formally, by explicitly computing the functional determinant of the $\phi$ field in the presence of a background constant $A_{\tau}$. To do this, we denote the eigenvalues of the relevant differential operator by $\lam_{n,\vec{p}}$. They satisfy the following equation: 
\be
(-D^{\mu} D_{\mu} + m^2) \phi = \lam_{n,\vec{p}} \phi
\ee
and are labeled by an integer $n \in \mathbb{Z}$ describing the dependence on Euclidean time and a continuous 3-vector momentum $\vec{p}$. Explicitly, they are:
\be
\lam_n = \le(\frac{2\pi n}{\beta} - q A_{\tau}\ri)^2 + \om_{\vec{p}}^2 \qquad \om_{\vec{p}}^2 = \vec{p}^2 + m^2
\ee
and thus we would like to compute the following determinant:
\be
\sL_{\mathrm{eff}}[A_{\tau}] = \log \det \le[(-D^{\mu} D_{\mu} + m^2)\ri] = T \sum_{n \in \mathbb{Z}} \int \frac{d^3p}{(2\pi)^3} \log \le(\le(\frac{2\pi n}{\beta} - q A_{\tau}\ri)^2 + \om_{\vec{p}}^2\ri)
\ee 
The sum over $n$ is divergent, but this UV divergence is unrelated to the thermal physics we are studying. We can isolate this divergence by taking a derivative with respect to $\om_{\vec{p}}$ to find
\be
\frac{1}{2\om_{\vec{p}}} \frac{d}{d\om_{\vec{p}}} \sum_{n} \log \le(\le(\frac{2\pi n}{\beta} - q A_{\tau}\ri)^2 + \om_{\vec{p}}^2\ri) = \sum_{n} \le[\le(\frac{2\pi n}{\beta} - q A_{\tau}\ri)^2 + \om_{\vec{p}}^2\ri]^{-1}
\ee
Performing the sum over $n$ and integrating over $\om_{\vec{p}}$ we find that up to an overall (divergent) $A_{\tau}$-independent additive constant we have
\be
\log \det \le[(-D^{\mu} D_{\mu} + m^2)\ri] = T \int \frac{d^3p}{(2\pi)^3} \log \le(\sinh\le(\ha(\beta \om_{\vec{p}} + i q A_{\tau}\beta)\ri) \sinh\le(\ha(\beta\om_{\vec{p}} - i q A_{\tau}\beta)\ri)\ri) 
\ee
Let us now work in the limit where the particles are very massive -- i.e. where most of the contribution to the energy comes from the rest mass of the particle $\om_{\vec{p}} \approx m$ -- and also where this mass is high compared to the temperature -- i.e. $\beta m \to \infty$. Then we can expand the action in powers of $e^{\frac{-\beta m}{2}}$ and find the leading $A_{\tau}$ dependence to be: 
\be
\sL_{\mathrm{eff}}[A_{\tau}] \approx T \int \frac{d^3p}{(2\pi)^3} \le(e^{-\beta m} \cos(q \beta A_{\tau}) + \cdots\ri)
\ee
Here we have taken $A_{\tau}$ to be a constant; thus this result is actually the leading term in a derivative expansion in powers of $\p_{i} A_{\tau}$, and the functional form as a function of $A_{\tau}$ is precisely that shown in \eqref{effacfiniteT1}. 

Finally, we can understand this from an even slicker ``first-quantized'' language, involving particle worldlines rather than quantum fields. Recall that the gauge-invariant degree of freedom involving $A_{\tau}$ is the holonomy of the gauge field around the thermal circle,
\be
\exp\le(iq\oint_{S^1} A_{\tau}\ri) \ . 
\ee
In the limit where $A_{\tau}$ is constant in $\tau$ the holonomy becomes $e^{iq \beta A_{\tau}}$. Particle worldlines -- i.e. the quanta of the $\phi$ field -- that wrap the circle will feel this holonomy (with the particle feeling $e^{i\beta q A_{\tau}}$ and the anti-particle feling $e^{-i\beta q A_{\tau}}$), and each contribution of this worldline will be weighted by $e^{-\beta m}$ (i.e. the exponential of the particle action, which is just the length of its trajectory in Euclidean time). Assembling the pieces -- and performing a dilute gas sum over many disconnected particle worldlines that are each localized in $\mathbb{R}^3$, but wrap the $S^1$ cycle -- we again find the answer above. Note that this is essentially equivalent to the mechanism of Polyakov confinement in 3d \cite{Polyakov:1976fu,Polyakov:1975rs}, with the role of the monopole instanton now being played by the wrapped electric worldline. 

Let us now understand the spatial components $A_i$; as it turns out, in the usual thermal equilibrium these components are essentially unaffected and they just inherit their bare kinetic term. (In the plasma literature, this is usually stated as the fact that the ``magnetic field is unscreened''). 

Assembling the pieces we find the following answer for the full effective action in 3d:
\be
S[A_{\tau}, A_i] = \beta \int d^3x \le(\frac{1}{2 e^2} (\p_{i} A_{\tau})^2 + \frac{1}{4 e^2} (\p_i A_j)(\p^i A^j) + c_1 e^{-\beta m} \cos (q A_{\tau}) + \cdots\ri)
\ee
(If we move away from very massive limit, then we can still expand the effective action in harmonics of $\cos(q A_{\tau})$, in which case you can view this as the leading term.) The 3d symmetry currents that we previously identified can now be understood in terms of these low-energy fields as:
\be
J^{ij} = \ep^{ijk} \p_{k} A_{\tau} \qquad j^{i} = \ep^{ijk} \p_{j} A_{k}
\ee
Note the somewhat surprising fact that in the phase just described, the $U(1)^{(0)}_m$ is {\it spontaneously broken}. As we first saw back in the motivational Section \ref{sec:motivation}, it is possible to dualize the 3d vector field into a compact scalar $\th$ in \eqref{dualmap}, in terms of which the action for $A_i$ becomes that of a massless scalar action for $\th$. 
$\th$ is the Goldstone mode of the symmetry breaking of $U(1)^{(0)}_m$. 

%\begin{tcolorbox}
%{\bf Exercise:} 
%Prove the statement above: i.e. show that the action of free 3d electromagnetism with gauge potential $A_i$ can be dualized in terms of a compact scalar $\psi$, whose 0-form shift symmetry is spontaneously broken.
%\end{tcolorbox} 

In the application to plasmas, this captures the fact that the magnetic field is unscreened. If you imagine probing the finite temperature medium with two probe magnetic monopoles, they will feel a Coulomb law attraction, mediated by a gapless long-range force. This force arises from the Goldstone mode. 

Note that due to the coupling to the finite-temperature charged matter, $A_{\tau}$ is gapped out -- and therefore probing the system with two probe electric charges will {\it not} result in a long-range force. This is precisely the physics of {\it Debye screening}: the electric field is screened by free charges in the medium. 

The situation can be summarized in the table below; the phase we have just described is what is normally called {\bf plasma}\footnote{It is somehow soothing after all these years to find that this sophisticated pattern of higher-form symmetry breaking formally justifies the zeal with which my high-school science textbook distinguished between the ``plasma'' phase and conventional liquids and gases. It also allows one to formally distinguish between things like $\vec{D}$ vs $\vec{E}$ and $\vec{B}$ vs $\vec{H}$ and so on; the reader who wants to exorcise these possibly unpleasant memories using the sledgehammer of 1-form symmetry is directed to Appendix B of \cite{Das:2023nwl}. }

\begin{table}[h]
\centering
\begin{tabular}{c|c|c}
& $U(1)^{(1)}_m$ & $U(1)^{(0)}_m$ \\
\hline
Plasma (normal phase) & Unbroken & Spontaneously broken \\
Superconducting & Unbroken & Unbroken \\ 
\end{tabular}
\end{table}

Here we have also included the pattern of symmetry breaking for the superconducting phase at finite $T$. This is essentially the same as in Table \ref{EMsym} at zero temperature. 

Note that the finite-temperature phase transition between the superconducting and plasma phase at finite temperature is just the breaking of an ordinary $U(1)$ 0-form symmetry in the 3 (non-compact) directions. In general this is the {\it XY universality class}. The application to finite-temperature superconductors is called {\it inverted XY criticality} \cite{Peskin:1977kp,dasgupta1981phase,kiometzis1994critical,herbut1996critical} as you will notice the interesting fact that the plasma (high temperature) phase has the symmetry spontaneously broken. This counterintuitive behavior arises ultimately from the origin of the finite temperature 0-form symmetry from the reduction of a 1-form symmetry on the thermal circle. The consequences of this pattern of symmetry breaking for a higher-form description of hydrodynamics was described in \cite{Armas:2018zbe,Armas:2018atq}. Discussion in language similar to that above can be found in \cite{Iqbal:2020lrt,Das:2023vls}. 

\subsection{Finite temperature: magnetohydrodynamics} 
Let us now think about the {\it real-time dynamics} of the finite-temperature QED plasma. This is the area that is the domain of {\it hydrodynamics}. 

\subsubsection{Lightning review of ordinary hydrodynamics} 

Let us briefly review the philosophy of hydrodynamics: it is an amazing fact that essentially any interacting system (quantum or otherwise) at finite temperature basically behaves at long distances and late times as a {\it fluid}, whose structure is dictated by the global symmetries and conserved charges of the system \cite{landau1987fluid}. 

To understand this, note that the excitations of most of the degrees of freedom in a fluid will decay away with a characteristic time-scale $\tau$ that is set by some microscopic scale (e.g. the mean free path $\tau \sim \frac{\ell_p}{v}$, or the temperature $\tau \sim T^{-1}$, etc.). However, conserved quantities cannot vanish arbitrarily fast: instead they must ``spread out'', and the speed of this process depends on the size of the excitation. Thus if one creates a lump of charge with some size $L$, the speed of its decay will decay as some inverse power of $L$ ($\tau \sim L^{-2}$ for the most generic diffusive processes). At late times only the conserved charges are left. 

In the modern formulation of hydrodynamics, this philosophy has been promoted to an algorithm: given a structure of global symmetries, one can turn a crank and a hydrodynamic theory pops out \cite{Kovtun:2012rj}, which can be expressed in terms of coarse grained variables. 

In the simplest case of a relativistic case with a conserved stress tensor
\be
\p_{\mu} T^{\mu\nu} = 0 \label{consT} 
\ee
the fluid degrees of freedom are a normalized fluid velocity $u^{\mu}(x)$, $u^2 = -1$, and a scalar temperature field $T(x)$. The microscopic stress tensor can be expressed in terms of these coarse-grained variables as
\be
T^{\mu\nu} = (\ep + p)u^{\mu} u^{\nu} + p g^{\mu\nu}
\ee
Here $\ep(T)$ and $p(T)$ are functions of the temperature through the {\it equation of state} of the fluid, and the conservation of the stress tensor becomes a relativistic Navier-Stokes equation for $u^{\mu}$. If we further consider the case of hydrodynamics with an ordinary $0$-form global symmetry then we also add a degree of freedom $n(x)$ for the local charge density and write
\be
T^{\mu\nu} = (\ep + p)u^{\mu} u^{\nu} + p g^{\mu\nu} \qquad j^{\mu} = n u^{\mu}
\ee
The conservation equation $\p_{\mu} j^{\mu} = 0$ provides the required dynamical equation of motion for $n(x)$.  

Finally, we stress that hydrodynamics is an {\it effective theory}, i.e. the expressions above can be systematically improved order by order in derivatives. Very important physics appears at first order in derivatives, which we discuss in the higher-form context below.  

\subsubsection{Magnetohydrodynamics}
Let us now apply this formalism to studying QED (with charged matter) at finite temperature. The hydrodynamic description of this model is normally called relativistic magnetohydrodynamics; it is conventionally studied by coupling the {\it gauged} electric charge current in the formalism above to Maxwell's equations in some form (see \cite{bellan2008fundamentals} for a textbook treatment and \cite{Hernandez:2017mch} for a modern discussion). 

We will not do this -- instead we will go straight to the true 1-form global symmetry. 

The conserved quantities are the regular stress-energy tensor $T^{\mu\nu}$ and the magnetic 2-form\footnote{In this section we use a different normalization for the current than in \eqref{Jdef}, to be closer to the fluid dynamics literature.} current $J^{\mu\nu} = \frac{1}{2} \ep^{\mu\nu\rho\sig}F_{\rho\sig}$. We now want to realize these objects in terms of some coarse-grained variables. This can be done by turning the hydrodynamic crank, but generalized to 1-form symmetries \cite{Grozdanov:2016tdf}. The idea of casting MHD as a fluid of strings predates the formal technology of 1-form symmetries, and was first discussed in \cite{Schubring:2014iwa}. 

Here we discuss only the results, referring you to \cite{Grozdanov:2016tdf} for the derivations. At zeroth order in derivatives we have:
\be
T^{\mu\nu}_{(0)} = (\ep + p)u^{\mu} u^{\nu} + p g^{\mu\nu} - \mu \rho h^{\mu} h^{\nu} \qquad J^{\mu\nu}_{(0)} = 2 \rho u^{[\mu} h^{\nu]} \label{MHD0thorder} 
\ee
Here $u^{\mu}$ is the regular timelike fluid velocity. $h^{\mu}$ is a spacelike vector which indicates the direction of the magnetic field; these satisfy
\be
u^2 = -1 \qquad h^2 = 1 \qquad u \cdot h = 0
\ee
$\ep$ and $p$ are the energy density and pressure as usual. From the expression for $J^{\mu\nu}$ we see that $\rho$ is the density of magnetic flux. Finally $\mu$ is a scalar which can be thought of as the chemical potential that is conjugate to $\rho$. There are two independent scalar quantities, which can conveniently be chosen to be $T$ and $\mu$, and there are the regular thermodynamic constraints between the scalars:
\be
\ep + p = Ts + \mu \rho \qquad dp = s dT + \rho d\mu
\ee
where $s$ is the entropy density.  

Note the term in $-\mu\rho h^{\mu} h^{\nu}$ -- this can be understood as the {\it tension} arising from the magnetic field lines. 

The conservation equations $\p_{\mu}T^{\mu\nu} = 0$ and $\p_{\mu} J^{\mu\nu} = 0$ now provide us with dynamics for the fluid degrees of freedom, which can be thought of as a kind of Navier-Stokes equation of $u^{\mu}$ and an equation telling us how the magnetic field $(\rho, h^{\mu})$ evolves. Note that we never need to use Maxwell's equations -- the equations close all by themselves, though one can reverse the logic and interpret the magnetic equation of motion as a kind of Maxwell's equation. 

We can now go to higher order in derivatives; the above is simply the first term in an expansion: 
\be
T^{\mu\nu} = T^{\mu\nu}_{(0)} + T^{\mu\nu}_{(1)} + \cdots \qquad J^{\mu\nu} = J^{\mu\nu}_{(0)} + J^{\mu\nu}_{(1)} + \cdots
\ee
Here for reasons of brevity we record only the first-order correction to $J^{\mu\nu}$, which takes the form: 
\be
J^{\mu\nu}_{(1)} = 2 m^{[\mu} h^{\nu]} + s^{\mu\nu}
\ee
where we have
\be
m^{\mu} = -2 r_{\perp} T \Delta^{\mu\beta} h^{\nu} \p_{\beta} \le(\frac{h_{\nu]}\mu}{T}\ri) \qquad s^{\mu\nu} = -2 \mu r_{\parallel} \Delta^{\mu\rho} \Delta^{\nu\sig} \p_{[\rho} h_{\sigma]}  \label{1order} 
\ee
We have introduced the projector $\Delta^{\al\beta} = g^{\al\beta} + u^{\al\beta} - h^{\al\beta}$, which annihilates both $u$ and $h$ and thus projects onto the subspace perpendicular to the magnetic field lines. We have introduced two new parameters $r_{\perp}$ and $r_{\parallel}$; these can be understood as {\it resistivities} that are parallel and perpendicular to the magnetic field lines. 

To determine the resistivities, we compute them from {\bf Kubo formulas}, which relate dissipative transport coefficients to retarded Green's functions of the current operator $J^{\mu\nu}$ in thermal equilibrium. If we place the background magnetic field in the $z$ direction, these Kubo formulas are: 
\be
r_{\parallel} = \lim_{\om \to 0} \frac{G^{xy,xy}_{JJ}(\om)}{-i\om} \qquad r_{\perp} = \lim_{\om \to 0} \frac{G^{xz,xz}_{JJ}(\om)}{-i\om}
\ee 

In \eqref{1order} there is a particular pattern of derivatives that is quite constrained; this arises from demanding consistency of the second law of thermodynamics, which guarantees that entropy always locally increases. Indeed once one works out the consequences of these terms, we find that there is {\it diffusion} of magnetic field lines. In the simplest case where we set the background magnetic field to zero, the two resistivities agree $r_{\perp} = r_{\parallel} = r$ and the dispersion relation takes the form \cite{Hofman:2017vwr}: 
\be
\om = -i \frac{r}{\chi} k^2 \qquad \chi = \frac{d\rho}{d\mu} \label{disprel} 
\ee 
where $\chi$ is the susceptibility of the 1-form charge.

This formalism is pleasing because it uses principles only of global symmetries and effective theory, and thus can be systematically improved and (in principle) applied also to situations where the normal textbook description of MHD breaks down, for example in situations of strong electromagnetic correlations. In principle such an approach might have real-world applications. The close connection to the global symmetry structure also allows the technology of holography to be brought to bear, an approach which has been very fruitful in understanding the structure of conventional hydrodynamics (see e.g. \cite{Son:2007vk} for a review). A sampling of work in developing the formalism and potential applications of such an approach includes \cite{Armas:2023tyx,Das:2022fho,Davison:2022vqh,Vardhan:2022wxz,Armas:2022wvb,Poovuttikul:2021fdi,Armas:2019sbe,Gralla:2018kif,Glorioso:2018kcp,Grozdanov:2018fic,Grozdanov:2018ewh,Grozdanov:2017kyl}. One concrete application is to the understanding of real-time lattice experiments, where lattice effects mean that the effective electromagnetic coupling $e$ can be of order $1$. See e.g. \cite{Das:2023nwl} for a recent numerical investigation which verified the dispersion relation \eqref{disprel}, and where this EFT approach allowed for an understanding of previous unexplained lattice results. 

\vspace{2cm} 
{\bf Conclusions:} In these lectures we had a whirlwind tour of several topics in generalized global symmetries. We have made no attempt at being complete and have left out many fascinating topics. A partial list of such omissions includes:
\begin{itemize}
\item Our discussion used very little mathematical formalism and thus did not at all address the construction of representation theory for non-invertible and higher-form global symmetries, a program which is currently undergoing rapid progress (see e.g. \cite{Bhardwaj:2022yxj,Bartsch:2023pzl,Bartsch:2023wvv,Bhardwaj:2023wzd,Bhardwaj:2023ayw}).
\item We only briefly discussed 't Hooft anomalies for conventional symmetries as a gateway to ABJ anomalies and their interpretation in terms of non-investible symmetries. There is clearly much more to say here: a review of anomalies in the conventional 0-form case can be found in \cite{Harvey:2005it}. Higher form symmetries can themselves have anomalies; indeed the electric and magnetic 1-form symmetries in free QED have a mixed anomaly, and a similar anomaly is present in any phase that spontaneously breaks a continuous $U(1)$ 0-form symmetry \cite{Delacretaz:2019brr}. Anomalies in higher-form symmetries can be used fruitfully to constrain the phase structure of gauge theories, see e.g. \cite{Gaiotto:2017yup,Gaiotto:2017tne,Gomis:2017ixy} for foundational early work. The formalism is reviewed in \cite{Bhardwaj:2023kri} and a review of applications can be found in \cite{Brennan:2023mmt}. 
\item {\it Higher groups} provide a mechanism by which symmetries of different forms can mix with each other and fuse into a larger structure; they are characterized by structure constants which are invariant under RG and can also be used to understand phases quantum field theory. Foundational work includes \cite{Bhardwaj:2017xup,Benini:2018reh,Cordova:2018cvg}, and developments are reviewed in \cite{Bhardwaj:2023kri}. 
\end{itemize} 

Nevertheless, a concrete lesson is that thinking about global symmetries as defining {\bf topological objects} results in many insights that are both profound and useful. A more philosophical lesson is that many interesting systems in nature involve the existence of extended objects (magnetic field lines, flux tubes, domain walls, etc.), and the community is currently in the process of developing a formalism that allows us to reformulate quantum field theory in terms of the dynamics of these objects, rather than in terms of local fields (such as gauge fields) who represent extended objects indirectly (through gauge redundancies). 

It remains to be seen where this line of reasoning will lead. It is left as a final exercise for the readers of these notes to figure this out. 

\begin{tcolorbox}
{\bf Exercise:} 
Figure this out.
\end{tcolorbox} 

\newpage
{\bf Acknowledgments:} I want to thank the students and organizers of the TPI School at Jena for their enthusiastic participation and insightful questions. I am also lucky to have many wonderful collaborators and friends who have taught me most of the things I've discussed here. I want to particularly thank John McGreevy for many delightful conversations about The Correctness of Landau. I'm also very grateful to Aleksey Cherman, Arpit Das, Iñaki Garcia Etxebarria, Adrien Florio, Saso Grozdanov, Diego Hofman, Theo Jacobson, Napat Poovuttikul, Tin Sulejmanpasic and David Tong. I am supported in part by the STFC under grant number ST/T000708/1. 

\bibliographystyle{utphys}
\bibliography{all}
\end{document}